\documentclass[usenatbib]{mnras}

\pdfoutput=1

\usepackage{natbib}
\usepackage{float}
\usepackage{graphicx}   
\usepackage{amsmath}    
\usepackage{amssymb}    
\usepackage{threeparttable}
\usepackage{lipsum}
\usepackage{rotating,times,graphicx,latexsym}
\usepackage[T1]{fontenc}
\usepackage{ae,aecompl}

\newcommand\be{\begin{equation}}
\newcommand\en{\end{equation}}

\newcommand{\eg}    {\textrm e.g.~}
\newcommand{\ie}    {\textrm i.e.~}

\newcommand{\msun}{\mbox{M$_{\sun}$}}
\newcommand{\lsun}{\mbox{$L_{\sun}$}}

\newcommand{\arcs}{\hbox{arcsec}~}
\newcommand{\arcm}{\hbox{arcmin}~}

\newcommand{\degree}{\mbox{$^{\circ}$}}

\newcommand{\ak}{\mbox{$A_{\it K}$~}}

\newcommand{\hii}{\mbox{H~{\sc ii}~}}

\newcommand{\uchii}{\mbox{UCH~{\sc ii}}}

\newcommand{\tco}{\hbox{$^{13}$CO}~}
\newcommand{\twco}{\hbox{$^{12}$CO}~}
\newcommand{\amo}{NH$_3$}

\newcommand{\kms}{\hbox{km~s$^{-1}$}}
\newcommand{\cms}{\hbox{cm$^{-2}$~}}

\newcommand{\vlsr}{\hbox{V$_{\rm LSR}$}}

\newcommand{\htt}{\hbox{H$_{2}$}~}

\title[YSO jets in the Galactic Plane]{YSO jets in the Galactic Plane from
UWISH2:\\ V - Jets and Outflows in M17}

\author[Samal et al.]
{M. R. Samal$^{1}$\thanks{E-mail:manash.samal3@gmail.com}, W. P. Chen$^{1}$, M. Takami$^{2}$, J. Jose$^{3,4}$, D. Froebrich$^{5}$  \\
$^1$Graduate Institute of Astronomy, National Central University 300, Jhongli City, Taoyuan County - 32001, Taiwan\\
$^2$Institute of Astronomy and Astrophysics, Academia Sinica, PO Box 23-141, Taipei 10617, Taiwan\\
$^3$Indian Institute of Science Education and Research, Rami Reddy Nagar, Karakambadi Road, Mangalam (P.O.) Tirupati  517507, India\\
$^4$Kavli Institute for Astronomy and Astrophysics, Peking University, Yi He Yuan Lu 5, Haidian Qu, Beijing 100871, China\\
$^5$Centre for Astrophysics \& Planetary Science, The University of Kent, Canterbury, Kent CT2 7NH, UK\\
}

\begin{document}
\date{}
\pagerange{\pageref{firstpage}--\pageref{lastpage}}
\pubyear{2018}
\maketitle

\begin{abstract}
Jets and outflows are the first signposts of stellar birth. Emission in the H$_2$ 1-0\,S(1) line at 2.122\,$\mu$m is a powerful tracer of shock excitation 
in these objects. Here we present the analysis of 2.0\,$\times$\,0.8 deg$^2$ data from the UK Widefield Infrared Survey for H$_2$ (UWISH2) in the 1-0\,S(1)
line to identify and characterize the outflows of the M17 complex. We uncover 48 probable outflows, of which, 93 per cent are new discoveries. We identified 
driving source candidates for 60 per cent of the outflows. Among the driving source candidate YSOs: 90 per cent are protostars and the reminder 10 per cent 
are Class\,II YSOs. Comparing with results from other surveys, we suggest that H$_2$ emission fades very quickly as the objects evolve from protostars to 
pre-main-sequence stars.  We fit SED models to 14 candidate outflow driving sources and conclude that the outflows of our sample are mostly driven by 
moderate-mass YSOs that are still actively accreting from their protoplanetary disc. We examined the spatial distribution of the outflows with the gas 
and dust distribution of the complex, and observed that the filamentary dark-cloud ``M17SWex'' located at the south-western side of the complex, is 
associated with a greater number of outflows. We find our results corroborate previous suggestions, that in the M17 complex, M17SWex is the most active
site of star formation. Several of our newly identified outflow candidates are excellent targets for follow up studies to better understand very early 
phase of protostellar evolution.
\end{abstract}


\section{Introduction} \label{int}
Jets and outflows are common signatures of stellar birth. They are thought to represent two aspects of
the same mass-loss phenomenon responsible for  removal of angular momentum from the star-disc system,
allowing accretion to proceed and the star to grow. Observationally, a strong correlation between mass accretion and mass-loss has been observed in
young stars \cite[\eg][]{hart98}, favoring the fact that outflows indeed play a significant role in the growth and evolution
of young stars. In this context, while much has been learned about the  evolved class II-III phases of pre-main sequence
(PMS) stars from  observations and modeling \citep[e.g.][]{cab07}, the evolution of class 0/I objects, where
the roles of the outflows are significant and stars get most of their mass, is less understood.
Similarly, another important related question concerning outflows is the driving mechanisms.
Models fall broadly into two categories of: from the interface between
the star's magnetosphere and disc, i.e. the ``X-wind model'' and from a wide range of disc radii, i.e. the
``disk-wind model'' \cite[details can be found in][]{ray07}. Understanding which one of the two models is dominant, requires
investigation of a large sample of outflow bearing young stars, as outflows show a wide variety of morphologies \citep[\eg][and references therein]{arce07}.
Therefore, detecting and characterizing outflows from the youngest protostellar sources holds the key towards the understanding of
the evolution of protostars as well as their launching mechanisms.

In this context, near-infrared emission at 2.12 $\mu$m (\htt $\nu$ = 1-0 S(1)) is one of the ideal tools to search for shock-excited outflows in terms of
jets and knots from young sources.
This line is an excellent tracer of hot (T $\sim$ 2000 K) and dense (n $\geq$ 10$^{3}$ cm$^{-3}$) gas excited by the fast shocks 
(10--100 \kms) caused by the interactions of jets  with the surrounding interstellar medium \citep[e.g.][]{stan02}.
 Moreover, for large-scale studies, 
outflow morphologies with near-infrared observations are more successful in identifying the outflow 
driving sources (particularly in the crowded region) compared to the molecular observations that typically lack 
the spatial resolution and depth to identify the fainter outflows.
Jets are believed to be the highly compressed ejecta of the accelerated entrained gas of the ambient medium.
The jets and  winds drive slower moving molecular lobes called bipolar molecular outflows.

Although many studies using \htt observations have focused on identification and characterization of outflow sources in nearby
clouds  \citep[e.g.][]{hoda07,davis07,davis09, kumar11,khan12,bally14,zhang15}, searches 
for outflow sources on galactic scales are still lacking. 
In the last decade, a few hundreds individual young stellar objects (YSOs) with outflows, i.e. so-called  ``Extended Green Objects (EGOs)'',
have been identified in the Galactic plane using the {\it Spitzer} 4.5 $\mu$m band observations \cite[e.g.][]{cygan08,cygan09} of the GLIMPSE 
survey. GLIMPSE is, however, a shallow survey--the faintest EGOs identified in GLIMPSE survey have surface brightnesses in 4.5 $\mu$m 
images  $\geq$ 4 MJy sr$^{-1}$.  In comparison, the majority of the diffuse green emission in NGC1333 (distance $\sim$ 250 pc)  has 4.5 $\mu$m 
surface brightness  $\leq$ 4 MJy sr$^{-1}$ \citep{gut08}. Thus GLIMPSE survey is  more sensitive to massive young stellar objects (MYSOs) outflows
in distant star-forming regions.

The advanced wide field sensitive infrared telescopes in recent
years have enabled the explorations of outflow candidates in the infrared waveband over galactic scales. 
In this regard, the
UKIRT Widefield Infrared Survey for \htt \citep[UWISH2;][]{fro11} conducted with United Kingdom
Infrared Telescope (UKIRT) at 2.12 $\mu$m (\htt $\nu$ = 1-0 S(1))  has opened a new avenue to
search for outflows from the YSOs \cite[\eg][]{ioa12a,ioa12b,fro16,mak18}, even possibly from low-mass YSOs \cite[e.g.][]{ioa12b}. 
We note the surface brightness limit of the UWISH2 survey in the \htt narrow-band filter is 300-2000 times better than the \htt 
emission strength of the GLIMPSE survey at 4.5 $\mu$m band \citep[][]{fro11}, thus in general a likely better tracer of 
outflows, however, we emphasize that in the region of high extinction the visibility of \htt emission 
can be poor compared to 4.5 $\mu$m emission. Moreover, in the context of protostar detection, in recent years, the instruments of the {\it Spitzer} Space Telescope and 
the {\it Herschel} Space Observatory have improved the resolution and sensitivity in the mid- and far-infrared domain 
where protostars emit the bulk of their energy. In fact,
many well-known class 0/I protostars have been detected between  3.6 and 8 $\mu$m with rising spectral energy distributions (SEDs)
between 24 to 70 $\mu$m  \citep[e.g.][]{enoch09,manoj13,dun15}. 
So combining {\it Spitzer} and {\it Herschel} data with the 2.12  $\mu$m observations, 
it is now possible to identify deeply embedded outflow bearing class 0/I protostars in a star-forming complex that 
are too faint or extincted  to be detected with the previous shorter-wavelength 
facilities (\eg 2MASS, UKIDSS, GLIMPSE surveys).

In this work, we make use of UWISH2, {\it Spitzer} and {\it Herschel} data to identify and characterize the outflow 
driving young protostars of the M17 cloud complex, which is at a moderate distance 
and currently producing young stars at a high-rate.
This work is organized as follows: 
Section 2 describes the  M17 complex. Section 3 discusses the 
various data sets and photometric catalogues. In Section 4, we discuss the identification of outflows and 
outflow driving candidate YSOs, and physical properties of the candidate YSOs by fitting the YSO models to their observed SEDs. Section 5 
discusses the general nature of outflows, comparison with other similar surveys and also presents a few interesting cases.  Section 6 
describes the star formation scenario of M17  with the aid of spatial distribution and 
correlation of the identified outflows with the gas and dust of the complex. Section 7 summarizes the various results obtained.

\begin{figure*}
\centering
\includegraphics[width=15cm]{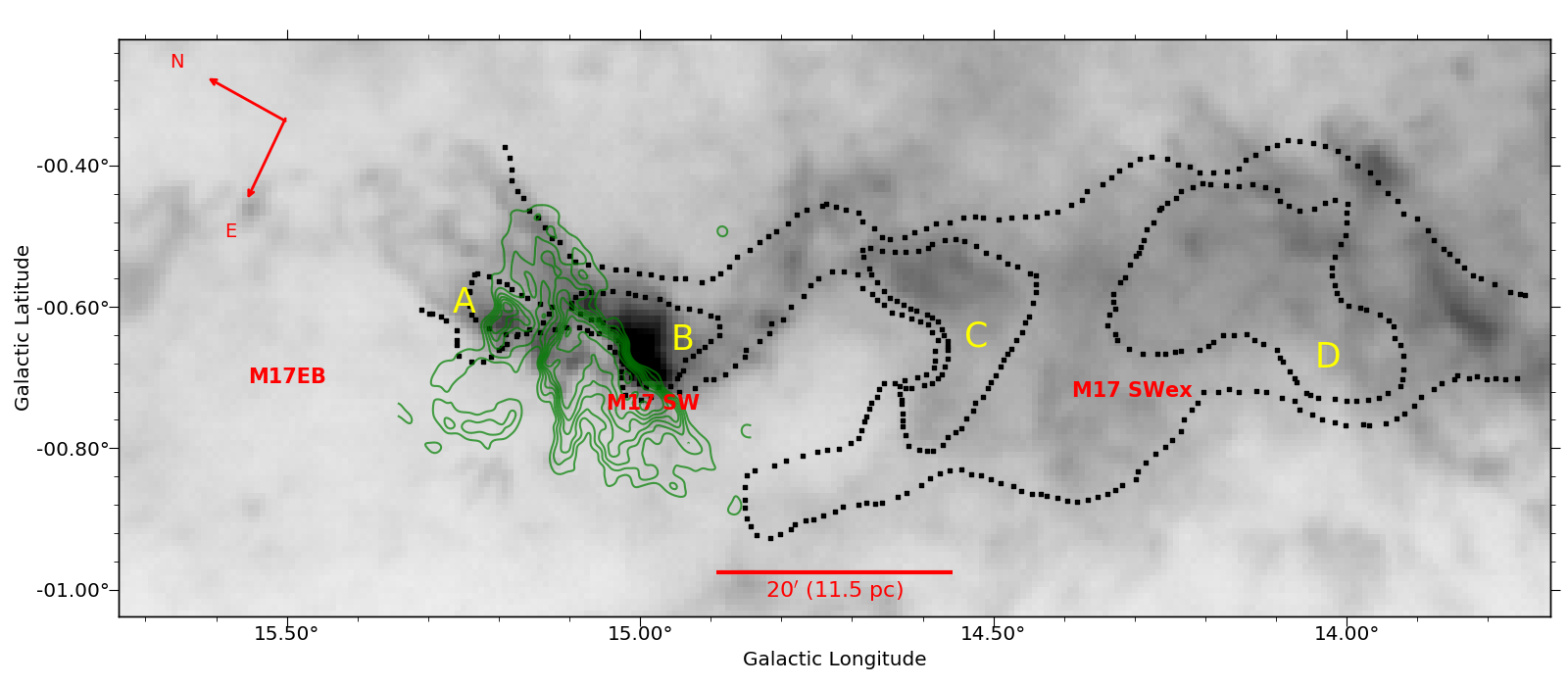}
\vspace{-0.3cm}
\caption{\twco map (grey-scale) of the M17 complex showing the distribution of molecular gas.
The map is obtained at a resolution of $\sim$ 46 arcsec with the Purple Mountain Observatory (PMO) 13.7-m telescope (Jiang et al. in preparation). 
The green contours correspond to the VLA 20 cm (at 0.05, 0.14, 0.24, 0.34, 0.43, 0.53, 0.63, 0.73, 0.83 and 0.93 Jy/Beam; 
resolution $\sim$ 6 arcsec) emission from \citet{hel06},
while the black contours represent the $^{13}$CO emission (at 5 K and 10 K; resolution $\sim$ 2.3 arcmin) from 
\citet{elm79}. 
The field size is $\sim$ 2.0 $\times$ 0.8 deg$^2$ (or 70 $\times$ 28 pc$^2$). The four dense fragments which 
has been designated by the letters A (in the NE) through D (in the extreme SW) are also marked.
}
\label{fig1}
\end{figure*}

\section {The M17 complex} \label{m17}

M17 ($l \sim 15^\circ.09,  b \sim -00^\circ.75$), located  at the north-east edge of one
of the largest giant molecular clouds (size $\sim$ 70 pc $\times$ 15 pc and mass $\geq$ 2 $\times$
10$^5$ \msun; \citet{elm79}) in the Sagittarius-Carina arm of our Galaxy, is a blister \hii region (also known as Omega Nebula,
W38, S45), and is the second brightest thermal radio
source in the sky after Orion. It is illuminated by the massive, 1--3 Myr
old \citep{jian02} stellar cluster NGC 6618, whose earliest O stars are
an O4-O4 binary  called `Kleinmann's Anonymous Star'.

The giant molecular cloud (GMC)  as observed in the low-resolution ($\sim$ 2.3 arcmin) $^{13}$CO, contains four
major fragments, namely A, B, C and D, each of mass $\geq$ 4 $\times$ 10$^4$ \msun~ \citep{elm79}. The local standard of rest velocity (\vlsr) 
and line width of these components are in the range of 20-22 \kms~ and 3.8-5.1 \kms, respectively. 
From the velocity structure of the $^{12}$CO and  $^{13}$CO gas,  \citet{elm79} suggests that most of the CO emission of the GMC comes 
from a single cloud of velocity around 20 \kms~ although the cloud has fragmented and there is a gradual change 
in velocity structure in the complex from northeast to southwest. The peak velocity increases from 20 $\pm$ 1 \kms~ in the northeast to about
22 $\pm$ 1 \kms~ to the extreme southwest. They argued that these differences in velocity pattern may be
related to the recent passage of a spiral density wave, which would have moved from northeast to southwest
in this part of the Galaxy.

Figure \ref{fig1} shows the overview of the complex, with all
the above components marked. It also shows the distribution of ionized gas associated with the \hii region. 
Among the fragments, the high-excitation temperature and compact
fragments, A and B, are located in the  northeast side of the complex very close to the \hii region, whereas the
low-excitation temperature and larger fragments, C and D, are located farther away from the
\hii region in its south-western direction. Among the fragments, the fragments B usually referred to as M17 SW \citep{thro83,gues88}, while
after the discovery of  flying dragon like dark cloud by {\it Spitzer} in 
the extended region of M17 SW (i.e. in the D component of the GMC), the D component is usually known as M17 SWex \citep{pov10,pov13}. 
The infra-red dark cloud (IRDC), G14.225-0.506, comprises the central region of M17 SWex, where dense filamentary ammonia gas (at \vlsr $\sim$20 \kms)
and hub systems have been observed \citep{bus13}.
Besides, \citet{pov07}, from the CO observations, detected a coherent shell-like structure in
the eastern side of the \hii region with local standard of rest velocity (\vlsr) $\sim$ 19 \kms. The shell is coincident with the
bubble walls of a cavity observed in the {\it Spitzer} images (Fig. 1). \cite{pov07}, named
it as the M17 extended bubble (M17 EB). 

The \vlsr~ and FWHM line width of the ionized gas associated to the \hii region is 18.6 \kms and 3.2 \kms, 
respectively \citep{jon86}.
Since the velocity of 
the various dense components \cite[]{elm79,bus13} are similar to the ionized gas velocity  of  M17, 
the bright \hii region and the dense fragments are part of the same cloud. The spectrophotometric distance for
NGC 6618 is estimated to be in the range
2.1--2.2 kpc \citep{chini80,hof08}, consistent with the kinematic distance $\sim$
2.2 kpc obtained from the radial velocity measurements of the ionized gas \citep{geo73,jon86}. Along the same line,
using  trigonometric parallax  of CH$_3$OH masers (\vlsr $\sim$ 23$\pm$ 3 $\kms$), the recent VLBI observations have
estimated a distance ${1.98}_{-0.12}^{+0.14}$ kpc \cite[]{wu14} to M17, which
we have adopted in this work. 

Star formation  towards M17 has been extensively studied by \citet{pov07}, \citet{pov10} and \citet{pov16}.
Their results suggest that M17 is currently producing stars at a rate $\geq$ 0.005 \msun~ yr$^{-1}$, $\sim$ four times 
the star formation rate of the Orion Nebula Cluster \cite[see][]{pov16}. Thus a potential site where one would expect 
a large number of protostars. In the present work, we  focus our study on an area $\sim$ 2.0 $\times$ 0.8 deg$^2$ of the complex,
primarily covering the dense clouds and the extended bubble.

\section {Observations and Data sets} \label{sec_obs}
We retrieved the narrow-band continuum subtracted \htt $\nu$ = 1-0 S(1) 2.12 $\mu$m images  from the
UWISH2 database\footnote{http://astro.kent.ac.uk/uwish2/} \citep{fro11}.
UWISH2  is an unbiased survey of the inner Galactic plane in the
\htt line at 2.12 $\mu$m, using the WFCAM camera at UKIRT. The survey covers $\sim$ 209 deg$^2$ along the inner Galactic Plane from
$l$ $\approx$ 357\degree~ to $l$ $\approx$ 65\degree~
and |$b$| $\le$ 1\degree.5.
WFCAM houses four Rockwell Hawaii-II (HgCdTe 2048 $\times$ 2048) arrays spaced by 94 per cent in the focal plane. For UWISH2 observations,
12 exposures were acquired at each telescope pointing, resulting in a total exposure time per pixel of 720 s. The typical full width half-maximum (FWHM) 
of the stellar point spread function (PSF) of the UWISH2 observations is 0.7 arcsec, and the typical 5$\sigma$ detection limit of 
point sources is $\sim$ 18 magnitude and the surface brightness  limit is $\sim$ 4.1 $\times$ 10$^{-19}$ W m$^{-2}$.
In order to perform the continuum subtraction of the narrow band images UWISH2 uses the UKIDSS data in the 
K-band (2.2 $\mu$m) taken with the same telescope and instrumental set-up as part of Galactic Plane Survey
\citep[GPS;][]{lucas08}. The GPS survey maps 1800 square degrees of the galactic plane (|b| $<$ 5\degree) in the J, H and K bandpasses 
with total exposure time per pixel of 80, 80 and 40 s, respectively, reaching 5$\sigma$ detection limit 
of J $\sim$ 19.8, H $\sim$ 19.0, and K $\sim$ 18.0 magnitude, respectively. The typical FWHM of the GPS survey is 0.9 arcsec.
We used the continuum subtracted \htt images to identify jets 
and knots in the M17 complex.

The M17 complex was observed  by the {\it Spitzer} Space Telescope as part of the Galactic Legacy Infrared Mid-Plane Survey Extraordinaire 
\citep[GLIMPSE;][]{ben03} and the Multiband Imaging Photometer GALactic plane survey  \citep[MIPSGAL;][]{car09b}. 
We obtained the Post Basic Calibrated Data (PBCD) images of the {\it Spitzer} Infrared Array Camera (IRAC) 
at 3.6, 4.5, 5.8 and 8.0 $\mu$m, and of the Multiband Imaging Photometer (MIPS) at 24.0 $\mu$m from the {\it Spitzer} 
Archive\footnote{http://sha.ipac.caltech.edu/applications/Spitzer/SHA/} to
search for  embedded YSOs and to examine the morphology of the inter-stellar medium (ISM)
at the location of jets and knots. For point sources, we used the UKDISS  $JHK$ \citep{lucas08}, 
GLIMPSE IRAC \citep{car09b} and MIPSGAL 24 $\mu$m \citep{gut15} point 
source catalogues available on the Vizier\footnote{http://vizier.u-strasbg.fr/viz-bin/VizieR-4} interface.
Note that, the 
3.6 and 4.5 $\mu$m bands are more sensitive to stellar photospheres than the 5.8, 8.0 and 24 $\mu$m bands, and 
the angular resolution of the images at IRAC bands is in the range 1.7--2.0 arcsec, whereas at  MIPS 24 $\mu$m band 
it is $\sim$ 6.0 arcsec.  

For this work, we also downloaded the 70 $\mu$m images from the {\it Herschel} Science Archive\footnote{ 
http://archives.esac.esa.int/hsa/whsa/}, observed as  part of the Hi-GAL survey \citep{moli10},
using the Photodetector Array Camera and Spectrometer (PACS). 
The angular resolution of the  Hi-GAL data at 70 $\mu$m  is $\sim$ 10 arcsec.  
To get point source fluxes, we performed photometry on the {\it Herschel} 70 $\mu$m 
image using the IRAF tasks {\it daofind} and {\it apphot} to extract the positions of the sources and 
to perform aperture photometry. We used an aperture radius of 12 arcsec, and inner and outer sky annulus
of 35 and 45 arcsec, respectively, and applied the aperture correction as documented
by the PACS team \citep{bal14}.
Since our target sources are embedded in strongly varied spatial structures, 
the variations in the background limit the photometric accuracy. We thus used different 
apertures and estimated that our photometry is accurate within 
10 to 15 per cent. 

The above point source catalogues are used to identify the YSOs in the vicinity of the jets/knots and construct their SEDs.
Note that the MIPSGAL 24 $\mu$m catalogue comes along with its GLIMPSE-IRAC counterparts. For making various 
color-color or color-magnitude plots between 3.6 to 70 $\mu$m (in search for YSOs close to the knots and jets), 
we matched  astrometic position of the MIPSGAL 24 $\mu$m catalogue with the astrometic position of the 70 $\mu$m sources 
using a position-matching tolerance of 6 arcsec\footnote{We note the astrometric accuracy of the MIPSGAL images is better than 1 arcsec, while
it is less that 2 arcsec for the Hi-Gal images}(the FWHM of the 24 $\mu$m data). For making the SED of the selected YSOs, we visually 
inspected the images and catalogues of the YSOs using {\it Aladin} software, then matched the already made 3.6 to 70 $\mu$m catalogue to
the  UKDISS catalogue using a matching radius of 2 \arcs (the FWHM of the GLIMPSE data) 
to obtain data points between 1.2 to 70 $\mu$m. In a few cases, where
there were more than one source within the matching radius, we considered the closest one as the best match.

In addition, we also exploit information from the 
following  major available  surveys in search for 
early stages of star formation such as cold cores/clumps  or SiO emission, 
in the vicinity of the jets/knots: 

\begin{enumerate}
\item  We used \citet{cse14}  catalogue 
of compact objects from the APEX Telescope Large Area Survey of the Galaxy (ATLASGAL) survey at  
870 $\mu$m (beam $\sim$ 19.2 arcsec,  sensitivity $\sim$  50--70 mJy/beam) for identifying clumps/cores.

\item  We used \cite{ros10} catalogue of compact objects  from the Bolocam Galactic Plane Survey \citep[BGPS;][]{agu11} at 1.1 mm (beam $\sim$ 33 arcsec, sensitivity $\sim$ 30--60 mJy/beam)
for identifying clumps/cores.

\item \citet{reid06} used the Submillimeter Common-User Bolometer Array (SCUBA)   on the 
James Clerk Maxwell Telescope (beam $\sim$ 15.4 arcsec, sensitivity $\sim$ 27 mJy/beam) to map an approximately 12 \arcm $\times$ 12 \arcm region 
around the M17 \hii region at 850 $\mu$m. 
We used their catalogue for our analysis.


\item 
SiO emission is a strong signpost of outflows from the very youngest, class 0/I, sources \citep[e.g.][]{Gibb04,cod07,taf15}, 
although other explanations are possible, such as low-velocity shocks caused by large-scale flow collisions during global collapse or 
by the dynamical interaction of two clouds \citep[e.g. see][and references therein]{lop16}.
Recently \cite{cse16} conducted a survey on massive clumps with the IRAM 30-m and APEX telescopes 
at the frequency of the SiO (2-1) and (5-4) transitions. 
In the present work, we used the catalogue of \cite{cse16} in search 
of SiO emission at the location of the MHOs, although this sample 
is only for clumps of mass greater than 650 \msun. Note that the spatial
position of the shock-excited SiO and 2.12 $\mu$m emission for a given outflow can be different as they
are sensitive to different physical conditions. 

\end{enumerate}

We note, owing to the different resolutions of the above mm-submm surveys, 
when we have multiple spatial positions for a given clump we considered 
the spatial position of the highest resolution 
as the better representation of its true position. We also note that the association of individual clumps/cores of the above surveys 
with the M17 complex required velocity information. However, 
it is worth mentioning that there are mainly two cloud components in the direction of M17 (particularly in the direction of C and D components), 
at \vlsr $\sim$ 20 and  40 \citep{lada76}; and as seen by \tco, the emission from the latter component is mainly distributed in the
latitude range -0.2\degree~ to 0.00\degree~ \citep[e.g.][]{umem17}, therefore unrelated to the area studied in this work. 
Moreover, from the work of \citet[][]{bus13} one can find that the most dense dust continuum clumps of M17SWex are 
positionally coinciding with the velocity of the ammonia gas at \vlsr $\sim$ 20$\pm$2 \kms. Since dense gas is more
related to star formation, we thus assume that contamination of other line of sight galactic dust 
clumps should be less in our studied area.

\section{Results}

\begin{table*}[t]
\scriptsize
\caption{\label{tab1} Properties of the MHOs}
\centering
\begin{threeparttable}[b]
\begin{tabular}{cccccccc}
\hline\hline
\multicolumn {1}{c} {ID} &\multicolumn {1}{c} {MHO number\tnote{1}} & \multicolumn {1}{c} {Lon.\tnote{2}} & \multicolumn{1}{c}  {Lat.\tnote{2}}  & 
\multicolumn{1}{c}{Length\tnote{3}} &\multicolumn{1}{c}{Source Type.\tnote{4}} & \multicolumn{1}{c}{Luminosity$^5$} & \multicolumn{1}{c}{Comments}\\

\multicolumn {1}{c} {} & \multicolumn {1}{c} {}   & \multicolumn{1}{c} {(degree)}  & \multicolumn{1}{c} {(degree)} & \multicolumn{1}{c} {(pc)}          &\multicolumn{1}{c}
{} 
&\multicolumn{1}{c}{($\lsun$)} &\multicolumn{1}{c}{} \\
\hline
01 & MHO 2308 & 13.78040 & -0.57323 & 0.22 & class 0/I &  27& bipolar, two opposite symmetric streamers\\
02 & MHO 2309 & 13.86546 & -0.54239 & -- & -- & & three interconnected  compact knots \\
03 & MHO 2310 & 13.90317 & -0.51448 & -- & cluster & & group of  compact knots \\
04 & MHO 2311 & 13.93151 & -0.49345 & -- & -- & &an isolated bright compact knot\\
05 & MHO 2312 & 13.96643 & -0.44227 & -- & -- & & two compact knots \\
06 & MHO 2313 & 14.07504 & -0.56333 & 0.20 & class 0/I & 27 &monopolar, an elongated patchy emission\\
07 & MHO 2314 & 14.08924 & -0.63322 & -- & -- & & an isolated compact knot with some diffuse nebulosity \\ 
08 & MHO 2315 & 14.09521 & -0.63386 & -- & -- & & an elongated knotty structure\\ 
09 & MHO 2316 & 14.11298 & -0.57489 & -- &  cluster & & a patch of diffuse emission\\ 
10 & MHO 2317 & 14.13259 & -0.52206 & 0.45 & class 0/I & 252&bipolar, a jet-shaped structure and two compact knots \\
11 & MHO 2318 & 14.14622 & -0.53602 & -- & -- & &an isolated compact knot\\
12 & MHO 2319 & 14.17600 & -0.67032 & -- & -- & & two aligned knot-like structures \\ 
13 & MHO 2320 & 14.17929 & -0.56165 & -- & class II & & monopolar, a patch of elongated faint emission\\
14 & MHO 2321 & 14.19136 & -0.50404 & -- & cluster & 13 &bipolar, chain of compact knots \\
15 & MHO 2322 & 14.19463 & -0.52274 & 0.22 & class  I/0 & 53& monopolar, two compact knots\\ 
16 & MHO 2323 & 14.20004 & -0.57689 & -- & -- & & two faint interconnected knots \\
17 & MHO 2324 & 14.21574 & -0.51510 & 0.82 & class 0/I & &bipolar, three compact jet-like structures along a line \\
18 & MHO 2325 & 14.21841 & -0.45140 & -- &  -- & & a bright head-tail like structure \\
19 & MHO 2326 & 14.22462 & -0.66556 & 0.09 & class 0/I & 34 & monopolar, an elongated head-tail like structure\\
20 & MHO 2327 & 14.22495 & -0.57625 & -- & -- & & three bright knots connected with some diffuse emission\\
21 & MHO 2328 & 14.24515 & -0.58420 & 0.30 & class 0/I & 22\tnote{*}& bipolar, two opposite asymmetric jets\\
22 & MHO 2329 & 14.24661 & -0.50227 & 0.30 & class 0/I & 53\tnote{*} &monopolar, a jet-like structure\\
23 & MHO 2330 & 14.27437 & -0.53161 & -- & --- & -- & two compact knots \\ 
24 & MHO 2331 & 14.27447 & -0.57511 & 0.78 & class 0/I & 19\tnote{*} &bipolar, chain of elongated faint emission\\
25 & MHO 2332 & 14.28056 & -0.49345 & -- & -- & & a bright compact knot\\
26 & MHO 2333 & 14.31092 & -0.59261 & 0.73 & class II & 286 & bipolar, chain of knots,  bend morphology\\
27 & MHO 2334 & 14.31447 & -0.53895 & -- & -- & & chain of elongated faint emission\\
28 & MHO 2335 & 14.32700 & -0.53257 & 0.55 & core & & monopolar, an elongated faint emission\\
29 & MHO 2336 & 14.33115 & -0.75052 &  -- & --  & -- & two bow-shock shaped structures\\
30 & MHO 2306 & 14.33145 & -0.64355 & 0.45 &  class 0/I & 4827\tnote{*}& bipolar, two bright elongated knots \\
31 & MHO 2337 & 14.35227 & -0.58742 & 0.07 & class 0/I & 4\tnote{*} & bipolar, two opposite symmetric jets  \\ 
32 & MHO 2338 & 14.36143 & -0.63861 & 0.41 & core & & monopolar, a compact knot with a faint streamer\\
33 & MHO 2339 & 14.36161 & -0.48936 & -- & -- & & a bright knot with some patchy emission\\
34 & MHO 2340 & 14.38114 & -0.68939 & -- & -- & & elongated faint diffuse emission\\
35 & MHO 2341 & 14.44860 & -0.56711 & 0.38 & core & &bipolar, two opposite streamer-like structures \\
36 & MHO 2342 & 14.60962 & -0.52421 & -- & -- & & a faint elongated haid-tail like structure\\
37 & MHO 2343 & 14.61739 & -0.60949 & 0.28 & class 0/I & & bipolar, compact knots with some patchy emission\\
38 & MHO 2344 & 14.63169 & -0.57720 & 0.43 & class 0/I & 715 & monopolar, multiple elongated knots \\
39 & MHO 2345 & 14.77681 & -0.48768 & 0.56 & class 0/I & 147 &bipolar, chain of patchy knots along a line \\ 
40 & MHO 2346 & 14.77861 & -0.33277 & 0.19 & class 0/I & 827 & monopolar, an elongated continuous flow \\ 
41 & MHO 2307 & 14.85161 & -0.99178 & 0.19 & class 0/I &  92 &  bipolar, compact bright knots \\
42 & MHO 2347 & 14.85168 & -0.98854 & 1.2 & class 0/I &  70 & bipolar, chain of knots \\
43 & MHO 2348 & 14.98216 & -0.67674 & 0.68 &  cluster & &monopolar, a bow-shaped structure with a faint tail\\
44 & MHO 2349 & 15.01505 & -0.63124 & -- & -- & & a compact bright knot with a diffuse tail\\
45 & MHO 2350 & 15.06827 & -0.61351 & 0.21 & core & & bipolar, two faint lobe-like structures\\
46 & MHO 2351 & 15.12872 & -0.49566 & 0.19 & class 0/I & 20 & bipolar, three compact knots \\
47 & MHO 2352 & 15.14967 & -0.61683 & 0.24 & core & & bipolar, two opposite bright lobes\\
48 & MHO 2353 & 15.25981 & -0.61653 & 0.6 & core &  & monopolar, two faint knots near to a core \\

\hline
\end{tabular}
\begin{tablenotes}
\item[1] The MHO numbers are assigned in the order of their right ascensions. Prior to this work, the MHOs 2306 and 2307 
have been identified by \citet{lee12} and \citet{carati15}, respectively. Thus they are appearing in the middle
of the table according to their right ascensions. 
\item[2] The coordinates are: (i) coordinates of the driving source if one found; (ii) coordinates of the peak intensity if 
the MHO constitutes an isolated knot; (iii) central
coordinates of the knots if the MHO constitutes multiple knots without a driving 
source candidate.
\item[3] Only for objects with a driving source we measured the apparent length. 
Quoted lengths are end-to-end for bipolar outflows or twice of the source-to-end for monopolar flows.
\item[4] We note, evolution 
of massive protostars differ significantly from the evolution of low-mass protostars \cite[e.g. ][]{hos09}, however, unlike the case 
of low-mass stars there is no observational evolutionary sequence that is firmly established for 
high-mass stars, thus in this work we have tentatively adopted the classification of the low-mass stars for the high-mass stars.
\item[5] Luminosities are either from SED models or using only 70 $\mu$m flux. Estimates based on the
later method are marked with asterisks. These luminosities are accurate within a factor of two.
\end{tablenotes}
\end{threeparttable}
\end{table*}

\begin{figure*}
\centering{
\includegraphics[trim={0.0cm 2.7cm 0.0cm 2.5cm},clip,height=5.0cm,width=13cm]{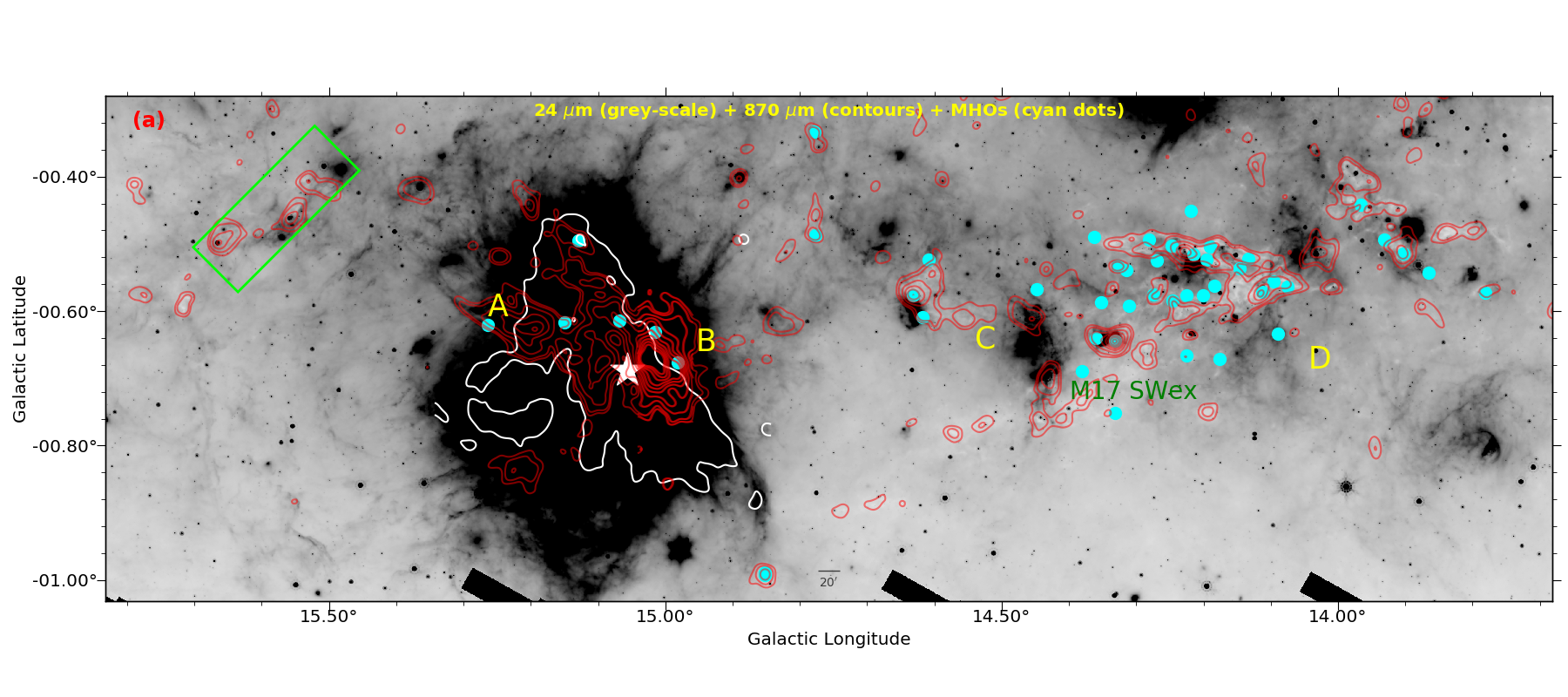}
\includegraphics[trim={0.0cm 1.0cm 0.0cm 2.5cm},clip,height=5.5cm,width=13cm]{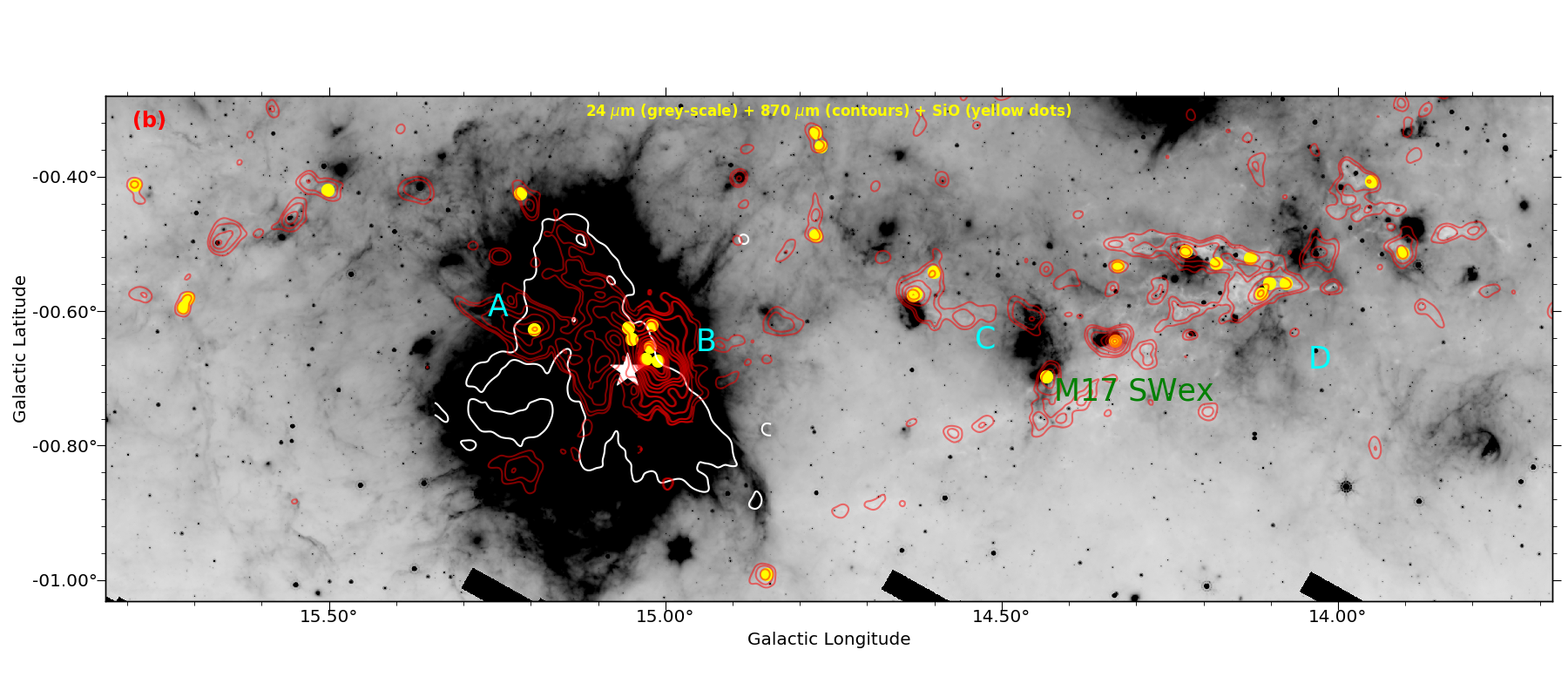}}
\vspace{-0.2cm}
\caption{Figure showing the distributions of MHO based outflows (top-panel) and SiO emissions (bottom-panel) in the M17 complex on the 24 $\mu$m background image. 
Red contours represent the 870 $\mu$m dust continuum emission from the ATLASGAL survey. 
Contour levels are at geometric progression in square root two, starting at 0.1 Jy/beam and 
ending at 7.5 Jy/beam. The outer extent of the main \hii region is shown in white contour ($\sim$ 0.05 Jy/beam at 21 cm).
The star symbol denotes the central location of the cluster NGC 6618. The 
green rectangle represents the area of the complex that is  not covered by UWISH2; it lies at the edges of 
UWISH2 tiles.  
}
\label{fig_ovpic}
\end{figure*}


\subsection{Identification of the outflows and outflow driving sources} \label{iden_mho}
Our first goal is to identify the shocked \htt features. To do so, 
we searched for  jets or knots or bow-shock features in the  continuum subtracted (\htt--K) images.
These sources tend to show up as enhanced emission over the background \htt--K image. However, 
identifying the true shock-excited features in the \htt--K images is not straight forward, as they can be confused
with several other artifacts. These include but are not limited to: (i) the fluorescence excited features by the UV 
field from a nearby massive star; (ii) the artifacts present in the residual images due to improper continuum subtraction; 
(iii) artifacts due to the sources of high proper motion or with excess emission.

One potential way to avoid  such  artifacts is to look at the multiwavelength data corresponding to
the \htt features. We therefore, checked the morphology and strength of the emissions at 1.2, 2.2, 4.5, and 8
$\mu$m images corresponding to each \htt emission. Artifacts due to improper continuum
subtraction or proper motion or artifacts of the bright stars can be  identified using 2.2 $\mu$m images 
\citep[e.g. see discussion in][]{ioa12a}.
Similarly, fluorescence excited \htt features can be picked up using  near-infrared colour-composite images and/or {\it Spitzer} 8 $\mu$m  images. Because fluorescence excited features appear brownish 
in the $JKH_2$  RGB images as  the $J$- and $K$-band continuum filters contain additional excited emission-lines \citep[e.g.][]{mak18}.
Similarly, 8 $\mu$m IRAC band contains the 7.7 and 8.8 $\mu$m emission features commonly attributed 
to polycyclic aromatic hydrocarbons (PAH) molecules \cite[][]{rea06}, and can be excited 
in the photo-dissociation-regions (PDRs) of a cloud by the  absorption of UV photons from nearby massive stars \cite[e.g.][]{pov07}. 
Thus, the  close resemblance of diffuse \htt and 8 $\mu$m emission at the peripheries
of the \hii regions or structures pointing toward the massive stars, are more likely representation of the
UV fluorescence excitation. 
In contrast to 8 $\mu$m, the  4.5 $\mu$m  band contains 
both \htt  and CO lines, that can become bright in the presence of shocked molecular gas, 
such as those expected from protostellar outflows \cite[e.g. see discussion in][]{crespo04,takami10,takami11}. Hence,  enhanced diffuse
emission  at 4.5 $\mu$m  at the location of \htt 
emission features, is a supportive indication of the presence of shocked outflow, although part of the 4.5 $\mu$m 
emission can be scattered continuum light from the embedded YSOs \citep{takami12}. In the literature, 
outflows identified based on the enhanced extended 4.5 $\mu$m emission are referred as ``EGOs''
\cite[\eg][]{cygan08} or ``Green Fuzzies'' \cite[\eg][]{debui10}, and are generally
identified by making IRAC colour composite images (using 8.0, 4.5 and 3.6 $\mu$m bands as red, green and blue colours, 
respectively)  as the 4.5 $\mu$m emission stands out against other IRAC band emissions. In this work, we have 
also used the {\it Spitzer} 4.5 $\mu$m and 8.0 $\mu$m band  as a tool by making IRAC colour-composite images  
to search for extra outflow components (if any) in the vicinity of 2.12 $\mu$m features (particularly in the highly 
extincted regions of a cloud such as dense clumps, as the extinction effect at 4.5 $\mu$m band is 
nearly half of that of the $K$-band) or to trace the PDRs  around
the massive stars.  

We also looked at the morphology of the \htt features to disentangle possible Planetary nebula \citep[e.g. see discussion in]
[]{ram17} from the wide-angle outflows.

Admittedly, the above criteria of identifying \htt features are somewhat subjective, and the possibility of missing 
faint and small \htt emission features exists. None the less,  comparing the number shocked \htt features identified by us 
with the automatically generated shocked \htt catalogue of the UWISH2 survey \citep{fro15}, we find that both the catalogues are 
in good agreement with each other. For example, both the catalogues have 93 per cent common sources. 
This ensures that the identified \htt features are highly reliable and affirmed that the false-positive rate in
the UWISH2 catalogue, if at all present, is likely to be less than 7 per cent. 
 
The  acronym `MHO' stands for molecular hydrogen emission-line object associated with the jets and outflows \citep{davis10}.
Several clusters of \htt shock features in the region can be clearly associated with coherent outflow, we therefore 
followed the procedure outlined in \citet{davis10},  i.e.  when it is possible to correlate multiple knots or jets to a single outflow,
we assigned them as a single MHO, otherwise, we considered each discrete jet or knot as an MHO. Briefly, we inspected large-scale 
\htt images in the search for possible large-scale flows. On the large-scale images, we looked for 
possible bow-shock features or aligned jets/knots. We then extended a line  tracing the middle of 
the bow-shocks or knot/jets in search for possible counter bow-shocks or jets/knots in the opposite symmetric axis.
For only a few cases, we observed a chain of jets/knots over parsec-scale dimensions, but the majority of the jets/knots are found to be
isolated or confined to only small spatial scale. We then based on the appearance/shape of the emission features, and the 
alignment of features with each other and/or the potential driving sources, we  assigned an MHO number to 
a jet/knot or  a chain of jets/knots or a group of jets/knots (explained in more detail below). With the above approach, 
we identified 48 likely outflows within 
our surveyed area. We list their positions in Table 1, while their distribution in the complex, 
and correlation with cold gas at 870 $\mu$m and shocked SiO emission are shown in Fig. \ref{fig_ovpic}. We discuss 
the general star-formation of the complex with the aid of these distributions  in Section \ref{over_sf}.
We note, as grouping weak emission features into an outflow is a complex problem, thus in this work, we treat each identified outflow
(particularly those without a driving source) as a candidate, pending verification through other shock 
tracers \cite[e.g.][]{plu13,zin15}, though most them will very likely turn into true outflows.

Here we describe our methodology of identifying YSOs and cores, and connecting them to the jets/knots.
The details concerning the classification of potential point sources into various YSO classes are given in Appendix A.
In summary, keeping in mind that the visibility of a point source at any given band is 
a strong function of the evolutionary status of the source itself, the extinction around its vicinity, 
and sensitivity of that particular band, 
we used several flux ratios between 3.6 to 70 $\mu$m to classify all those potential sources that are in the close vicinity of jets/knots into different
YSO classes (for details see Section \ref{iden_yso}). 
When possible we also used the available YSO catalogues from the literature. 
After identifying YSOs in the vicinity of jets/knots, we noticed that several  jets/knots are not associated with any YSO candidates, 
which led us to think that these jets/knots are possibly originating from molecular cores in their earliest evolutionary phases such as 
from first hydro- static cores \citep[e.g.][]{pezz12,ger15} or $70\,\mu {\rm{m}}$ dark cores \citep[e.g.][]{feng16,aso17}. It 
is also quite possible that: i) these jets/knots are part of large flows from distant sources; ii) 
the driving sources are still embedded in dense cores and are too faint to be detected in the {\it Spitzer} 
and {\it Herschel} bands (see Section 5.1 for further discussion).
To account for the sources that are either in the earliest evolutionary phases or deeply embedded in dense cores, we 
searched for early stages of star formation such as cold cores/clumps or infrared dark cloud fragments or SiO emission (see Section \ref{sec_obs})
in the vicinity of the jets/knots. After identifying YSOs and cores, in the next step we tried to connect them with the nearby jets/knots. 
Briefly, as the outflow symmetries can be different types, for example, 
as described in \citet{bally07}, S- and Z-shaped symmetries can occur if the
outflow axis changes over time due to precession induced by a companion or interactions with sibling
stars in a cluster, while C-shaped bend of outflow axis can occur due to the motion of 
surrounding gas or the motion of the outflow source itself, so identifying potential outflow driving 
sources is not trivial in the cases of misaligned outflows. Therefore, while connecting YSOs/cores with the jets/knots, 
we considered all those sources whose 
positions are compatible with the various jet shapes such as straight-, curved-, or S-shaped.
Along with the evolutionary status of the YSOs, we also used the indicators such as extended 4.5 $\mu$m and/or
shock tracer SiO emission  to approve or reject, whether or not 
any given source or core is responsible for the jets/knots observed 
in its vicinity. In a few rare cases, where we have two closeby sources, 
we gave higher priority to the younger and luminous YSOs, as 
luminous YSOs drive stronger outflows  \citep [e.g.][]{carati15,man16}.

\begin{figure*}
\centering
\includegraphics[width=11cm]{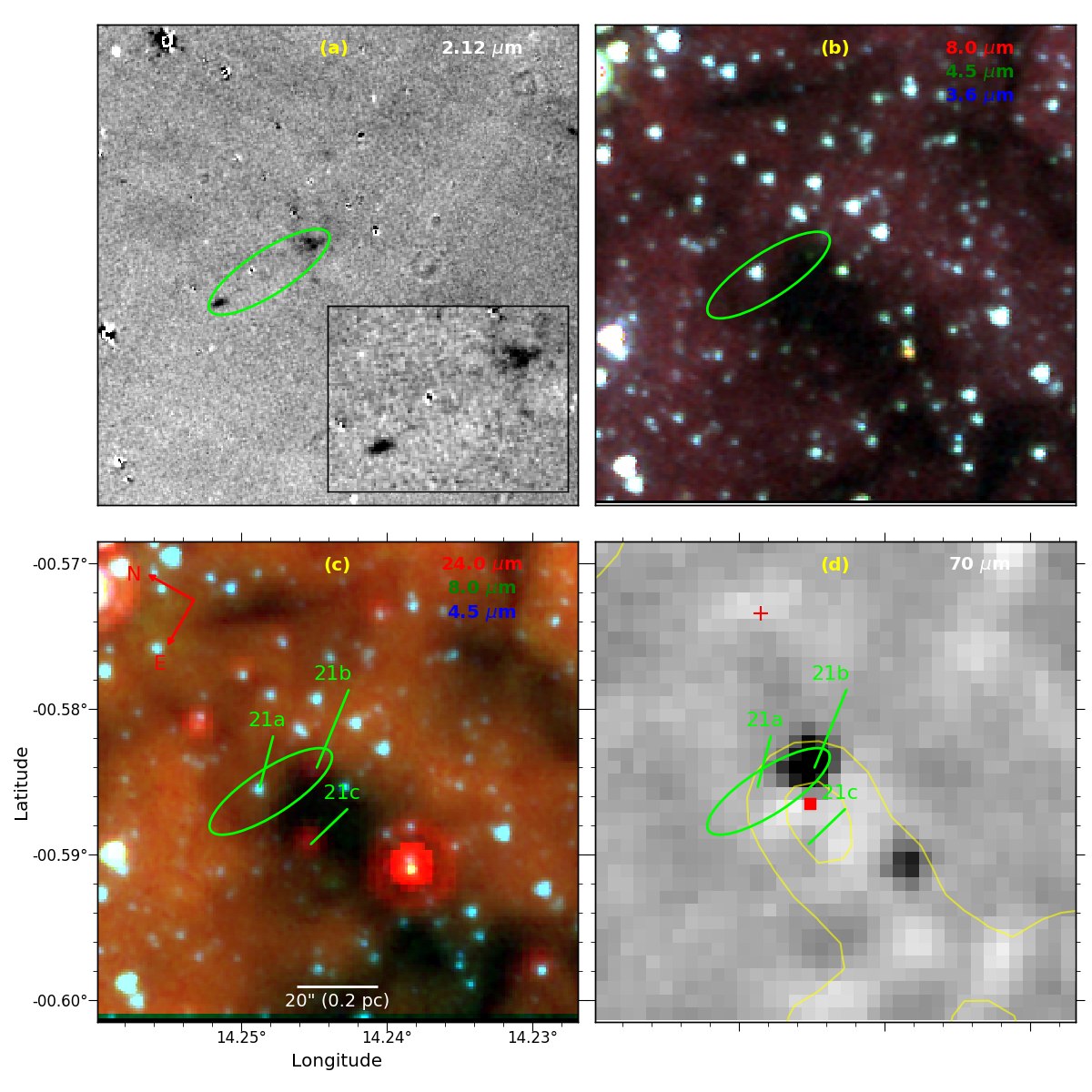}
\vspace{-0.3cm}
\caption{An example figure showing mutiwavelength view of ISM around an MHO. (a) The continuum subtracted \htt image, revealing the \htt jets.
The inset figure shows the zoomed view of the jets. 
(b) {\it Spitzer}-IRAC  colour-composite image (3.6 $\mu$m in blue, 4.5 $\mu$m in green and 8.0 $\mu$m in red); used to
search for enhanced 4.5 $\mu$m emission such as those found in EGOs. (c) 
{\it Spitzer}-IRAC/MIPS composite-colour image (4.5 $\mu$m in blue, 8 $\mu$m in green and 24 $\mu$m in red); used 
to unveil deeply embedded protostars. (d) The grey-scale unsharp masked 
70 $\mu$m image; used in search for early class 0 sources such as PACS Bright Red sources (see Section \ref{iden_yso}). 
Green contours show the distribution of  870 $\mu$m emission (from the ATLASGAL survey). 
The position of the outflow on the {\it Spitzer} and 70 $\mu$m images is shown with an ellipse and  the major axis of 
the ellipse indicates the likely flow direction.}
\label{fig_mho23}
\end{figure*}

\begin{figure*}
\centering
\includegraphics[width=11cm]{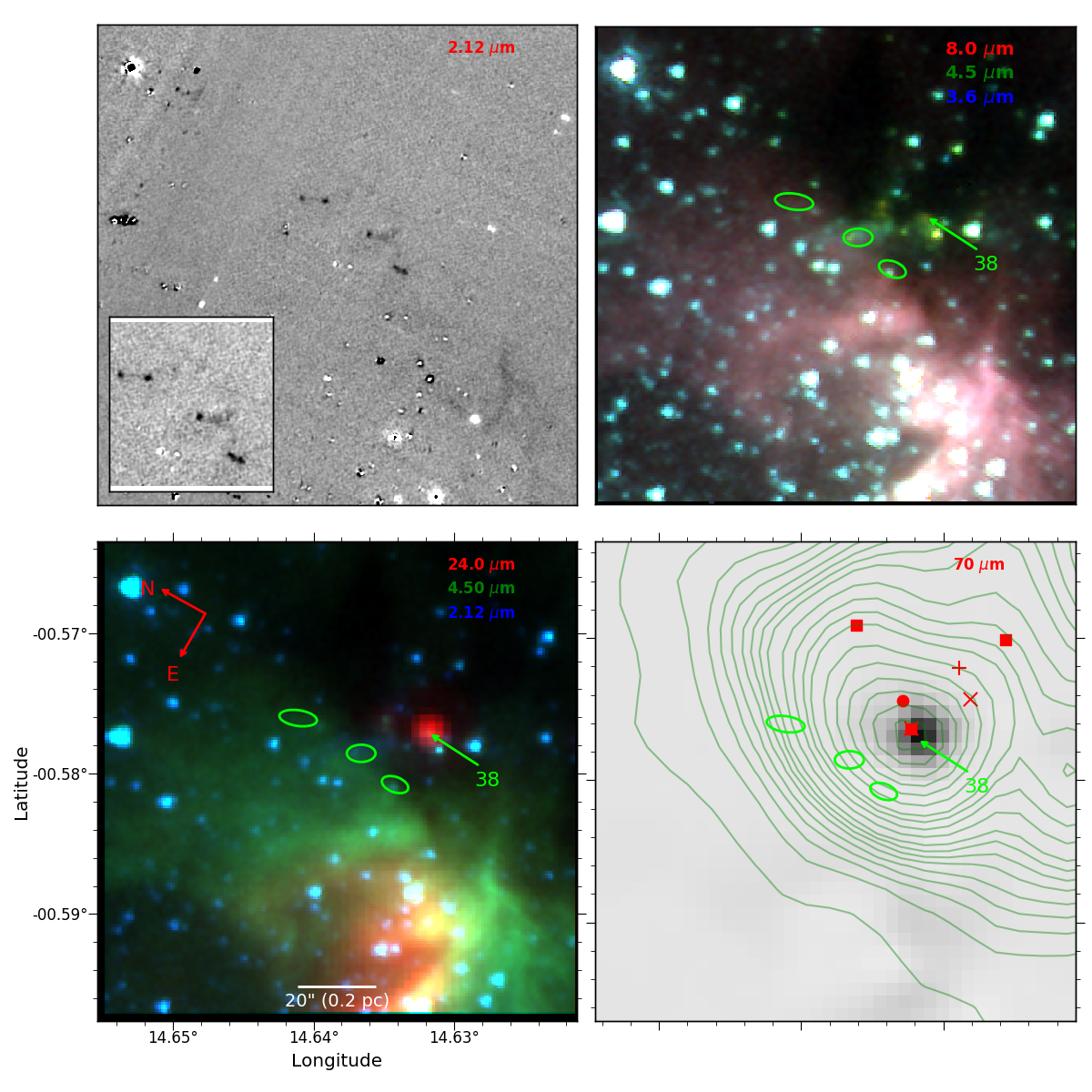}
\vspace{-0.3cm}
\caption{ An example figure showing  signs of star formation such as extended green emission, SiO maser ($\times$), cold clumps from the ATLASGAL 
($\blacksquare$) and  BGPS ($\bullet$) surveys and IRDC fragments ($+$) at the location of an outflow
consisting of three jet-like structures (rounded by oval shaped marks). In this work, we have considered that the jets are part of a single outflow 
whose origin lies in the clump (see text for details). 
All the images have the same meaning as in Fig. \ref{fig_mho23}. 
}
\label{fig_mho09}
\end{figure*}

Here we describe two complicated examples of our approach on the connection of the potential 
driving sources (cores or YSOs) with the jets/knots.
Figure \ref{fig_mho23} illustrates an  example of our approach on the search for potential driving sources around jets/knots. 
As can be seen, this region consists of  two elongated  \htt jets. Together they appear to form an east-west flow. 
Three potential  sources, visible in different bands between 3.6 to 70 $\mu$m 
(marked as 21a, 21b and 21c, in the figure), lies in the vicinity of the jets. Among the sources, the source 21a 
lies at the centre of jets and is visible in all the IRAC bands, the source 21b lies at one end of the MHO and  
is only visible in the 70 $\mu$m, while the source 21c 
lies slightly away from the flow axis, and is only visible at 
4.5 and 24 $\mu$m. Either of them can be the outflow driving source, although the possibility of 21c is less as it is 
not along the flow axis. The evolutionary status of the sources 21a and 21b are identified, respectively, in the Figs. \ref{iracmd}
and \ref{iracmd}c of Appendix A. In short, the source 21a is not a YSO as its location  coincides with the 
the zone of the field stars in the [5.8] versus [5.8] -- [8.0] diagram, 
it has no emission at 24/70 $\mu$m  and in the literature it is not an infrared excess YSO \citep[e.g.][]{pov10}. 
On the other hand,   
the source 21b  has the characteristics of an  early  class 0 source and it is bright in 70 $\mu$m. Although the  location of 21b is somewhat intriguing, but it could be due to
projection effect or anisotropy of the medium. It is worth mentioning, \cite{fro16}, in their study of the Cassiopeia and Auriga complex,
found that $\sim$ 20 per cent of the bipolar flows are asymmetric in nature with 
length ratio $<$0.5.  Since there are no other potential sources along the presumed 
flow axis of the jets (marked with an ellipse), thus 21b is the 
very likely driving source of the jets. 
Figure \ref{fig_mho09} depicts another example where three \htt jet-like features along with all the aforementioned signposts
of star-formation, including an EGO, have been found. 
As can be seen, the jet-like features lie  at the eastern outskirts of an ATLASGAL clump (shown in contours) and 
a bright 24 $\mu$m source (marked as ID 38)  lies at the centre of the clump. 
One can  see the middle jet connects well to the point source through the enhanced diffuse 
4.5 $\mu$m emission. The clump is also at the location of  SiO emissions. The column density in the direction of the clump is of the order of 
$\sim$ 10$^{23}$ \cms \citep[e.g.][]{cse16}, perhaps the possible reason why we do not see \htt emission in the vicinity of the source.
Deep high-resolution molecular observations would be needed to track the origin of these infrared jet-like structures \citep[e.g.][]{plu13,zin18}. 
None the less, it is worth mentioning, multiple wide-angle \htt bullets and jet-like structures have been  reported for 
a few cases \citep[e.g.][]{sah08,bal11}. The nature of such explosive outflows is still not well understood but it is believed that 
it might be related to dynamical decay of non-hierarchical system of stars or protostellar merger or close passage of two protostars
\citep[e.g. see discussion in][]{bal15,bal17,sah17}. In the present work, we tentatively consider that all the 
three jet-like features are part of a single outflow and are likely driven by the point source(s) embedded in the clump. The source 38 
being luminous and class-0 type, could be the dominant source responsible for the flow.

Figures \ref{fig_mho23} and  \ref{fig_mho09} are the illustration for two outflows. We followed a similar prescription for all the MHOs. The detailed discussion 
on the individual MHOs is given in  Appendix B, where we present multiwavelength large-scale ($\sim$ 1 $\times$ 1 pc$^2$ area)
images around the MHOs, prvide notes on individual objects and discuss their driving sources. 
Briefly, following the above approach, we associate 26 YSOs/cores and 4 clusters 
(i.e. the driving source is situated in a group of stars and we could not single it out) with 
the 48 MHO features. The spectral nature of the driving sources are also tabulated in Table 1. 
Out of 26 YSOs/cores, 6  are cores without point sources up to 70 $\mu$m, 18 are protostars (class 0/I YSOs) and 
2  are evolved sources (class II YSOs).


\subsection{Physical properties of the driving sources}
\subsubsection{SED modeling of the YSOs}  \label{sed_mod}

To get deep insight into the nature of the YSOs
identified in the present work, we modeled the observed  SEDs  
using the  models  of \citet{rob06,rob07}. The models used a 
Monte-Carlo based radiation transfer code of \citet{whi03a,whi03b} to follow photons 
emitted by the central star as they are scattered, absorbed and re-emitted throughout 
the disk and envelope system. The code uses a number of combinations of central star, disc, in-falling 
envelope and bipolar cavity, for a reasonably large parameter space. 
While other techniques (e.g. colour-colour diagrams or  
spectral indices) can be useful in identifying the evolutionary class, 
the SED models have the ability to infer physical information
about the young stars such as total luminosity, stellar age and mass and 
accretion rates, accounting the geometry of the disc and envelope.
However, interpreting SEDs using radiative transfer codes is
subject to degeneracies, which spatially-resolved multiwavelength
observations can overcome \cite[e.g. see][]{sam15}. 
Thus, we fit SED models to only those candidate YSOs for which we have constraints on the 70 $\mu$m flux along with
the fluxes at shorter wavebands between 1.2 to 24 $\mu$m. 

\begin{figure*}
\includegraphics[width=4.2cm] {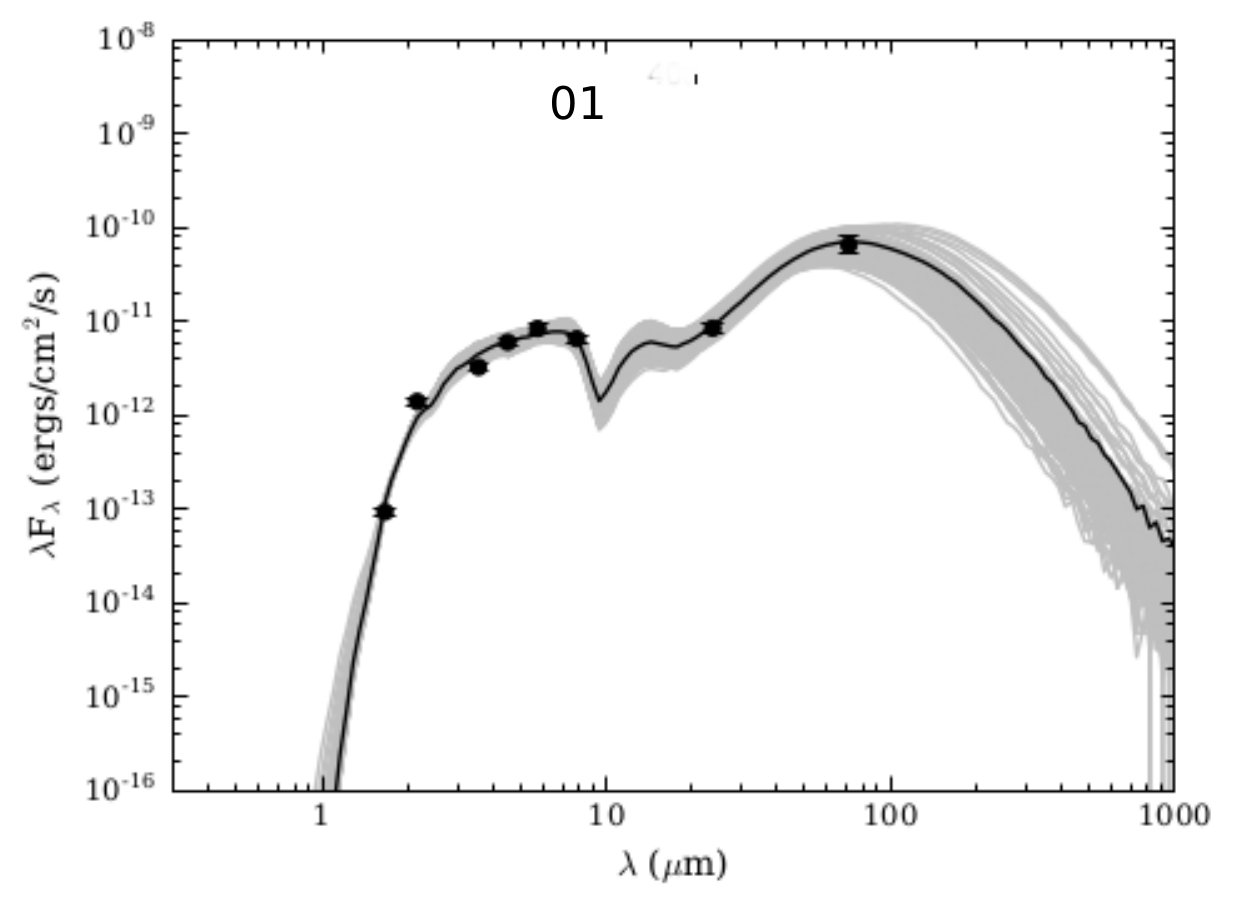}
\includegraphics[width=4.2cm] {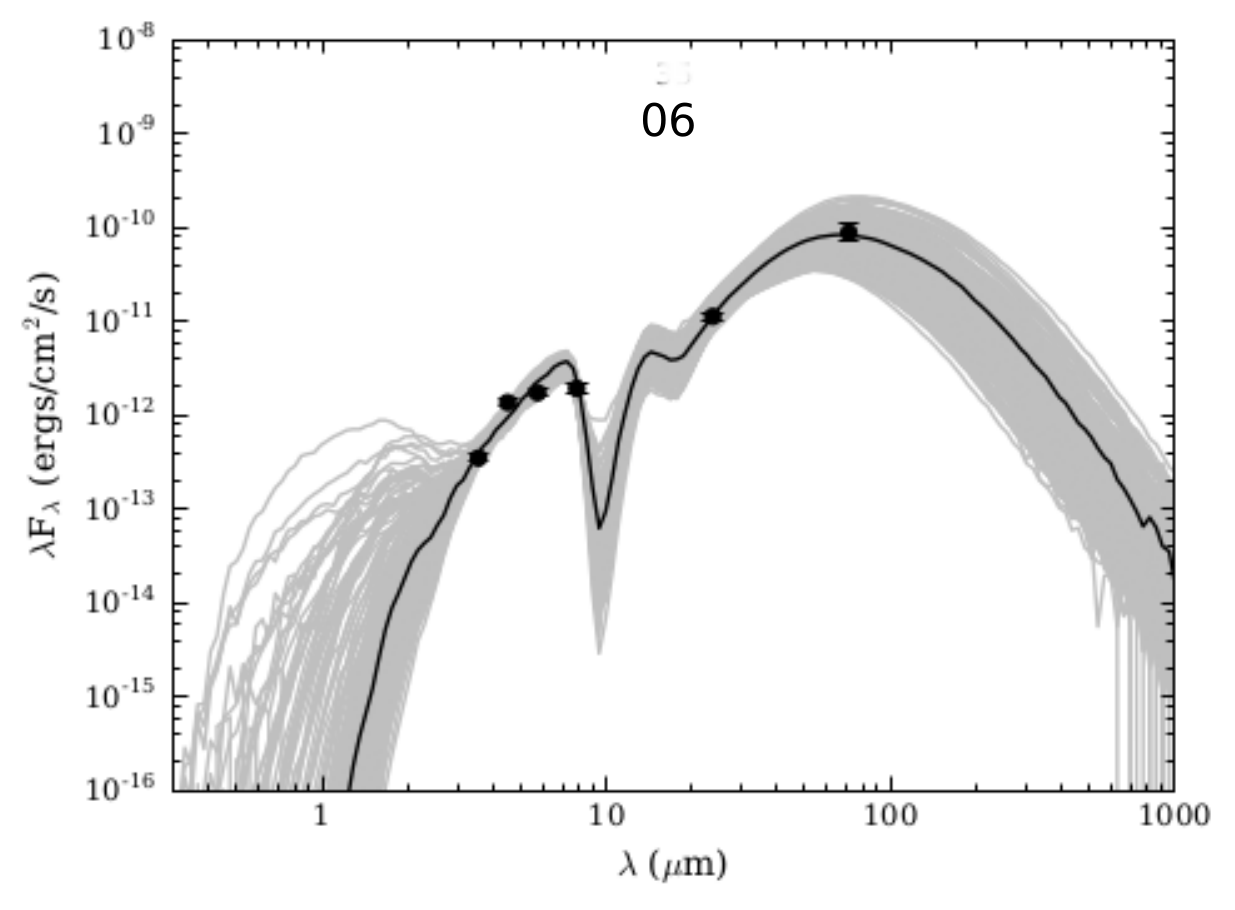}
\includegraphics[width=4.2cm] {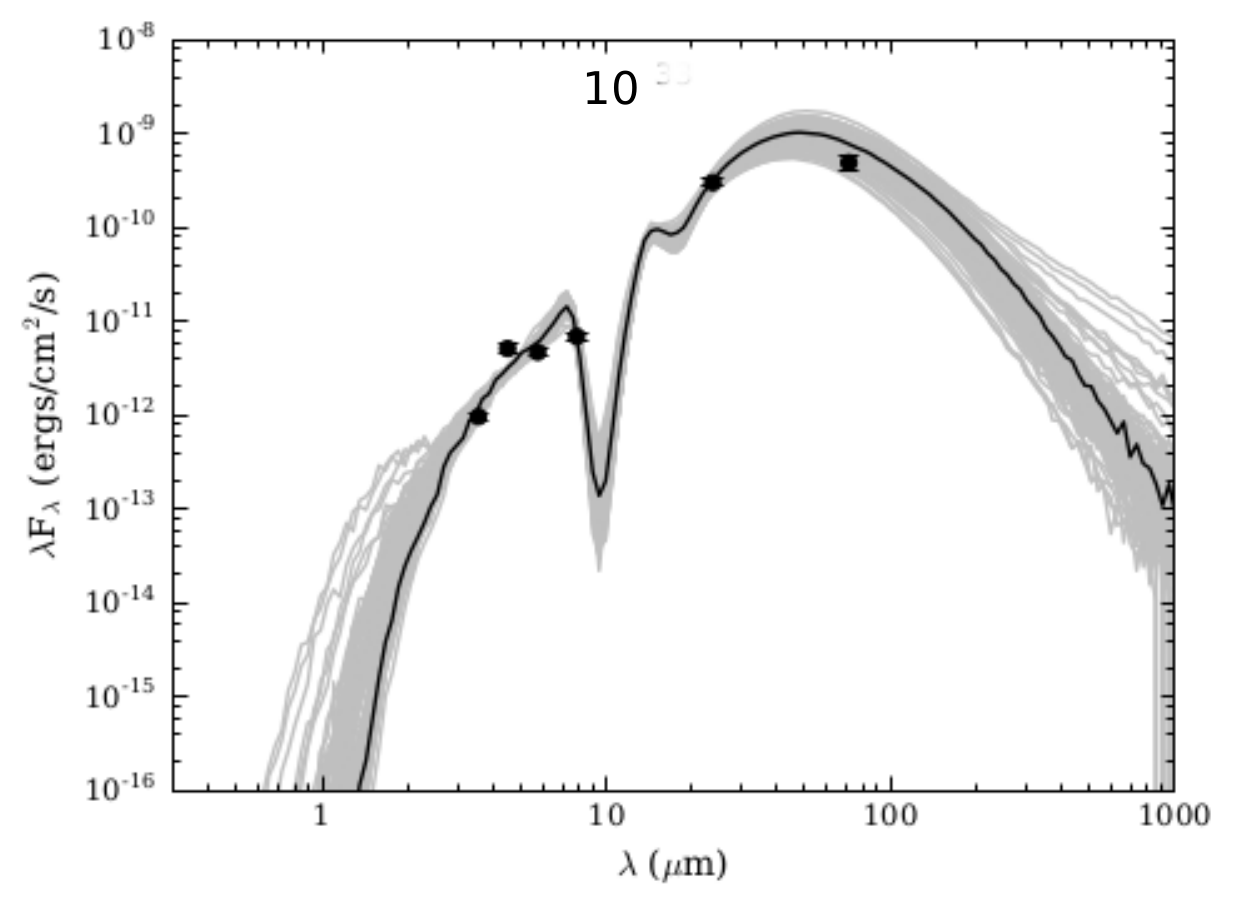}
\includegraphics[width=4.2cm] {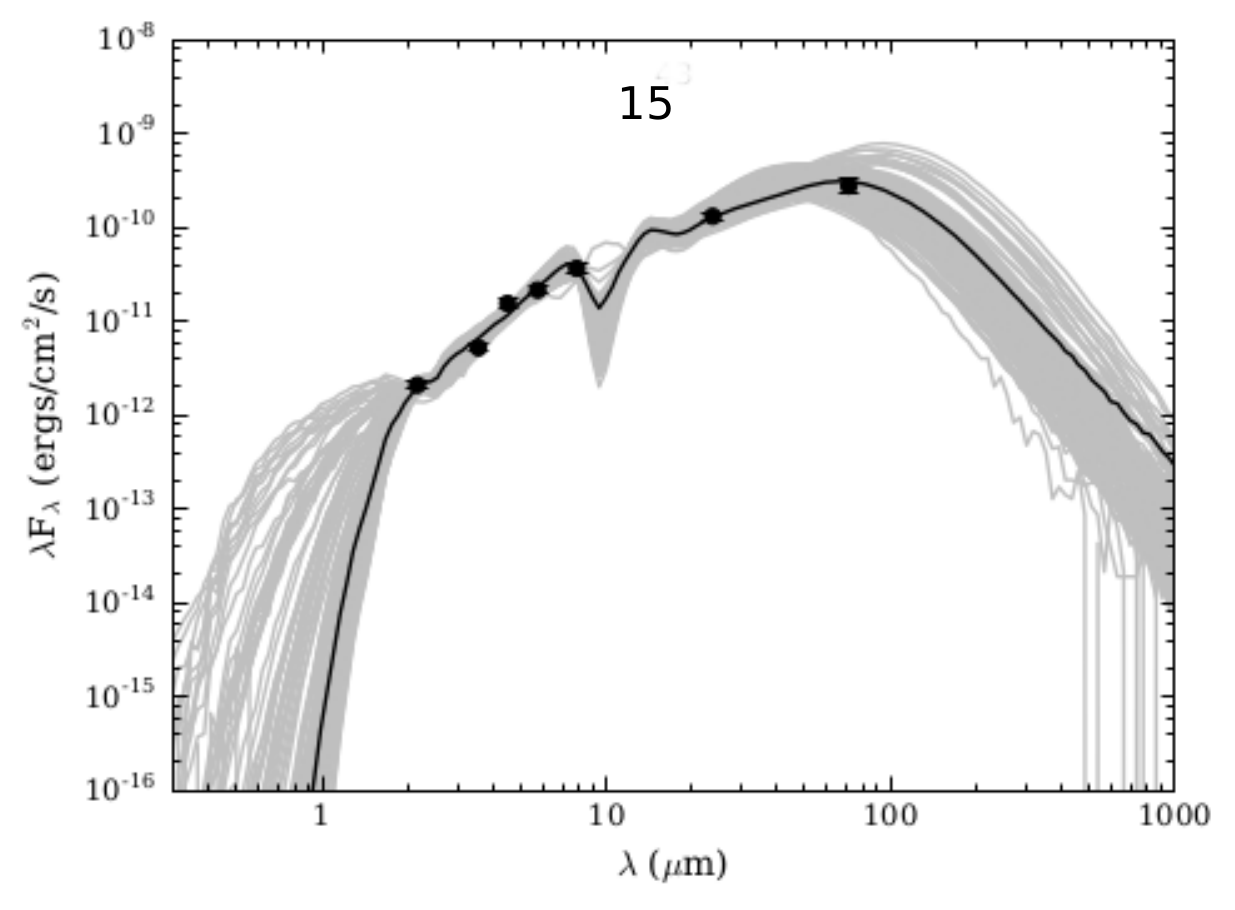}
\includegraphics[width=4.2cm] {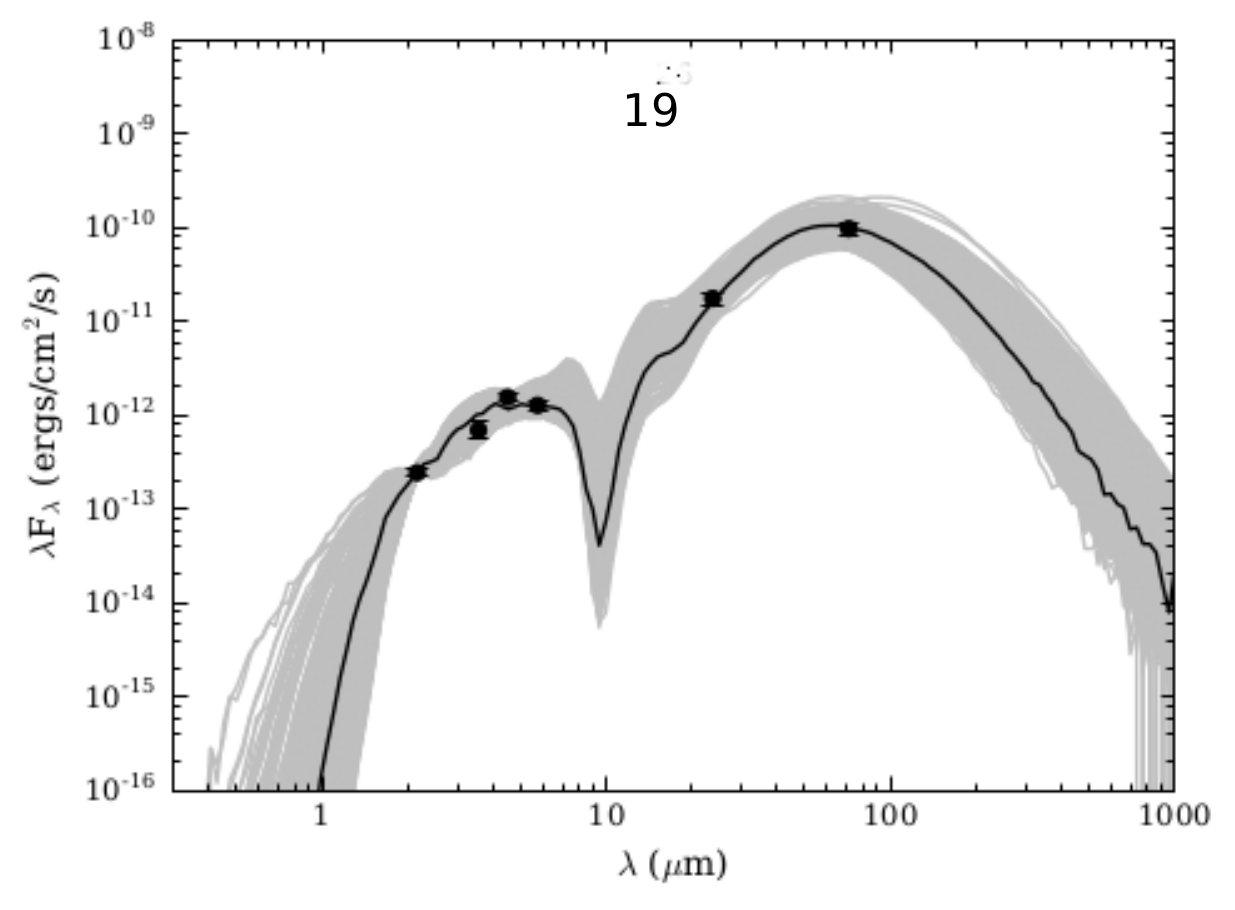}
\includegraphics[width=4.2cm] {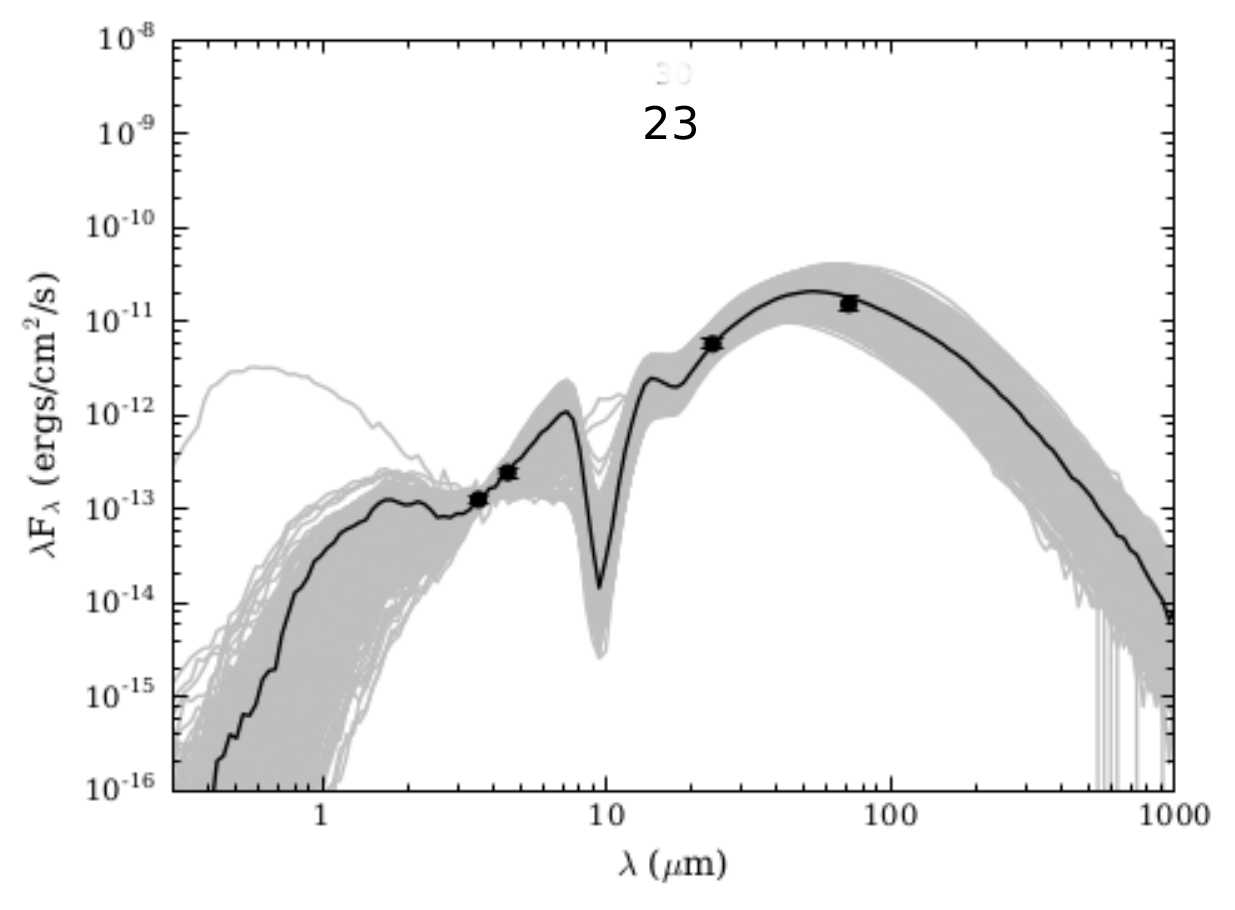}
\includegraphics[width=4.2cm] {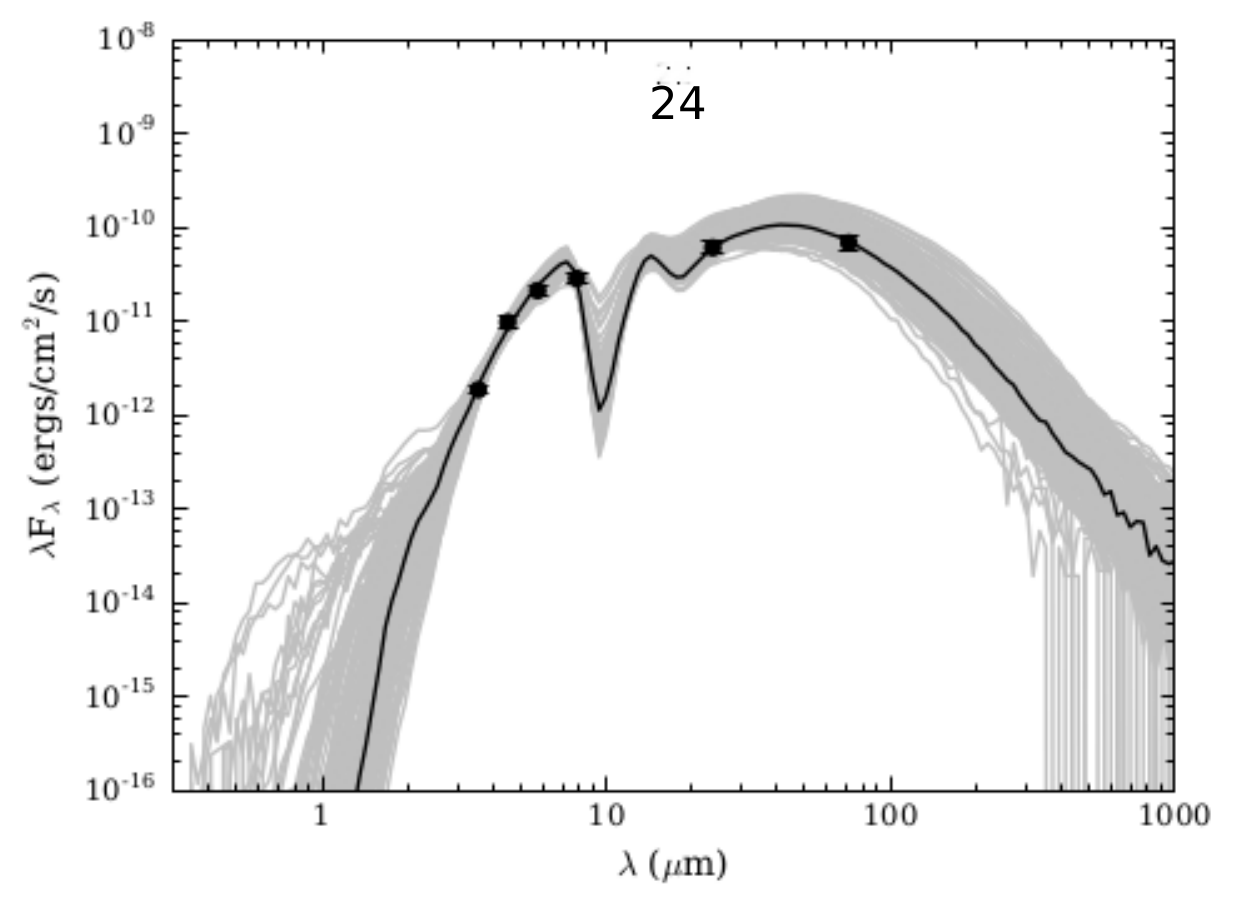}
\includegraphics[width=4.2cm] {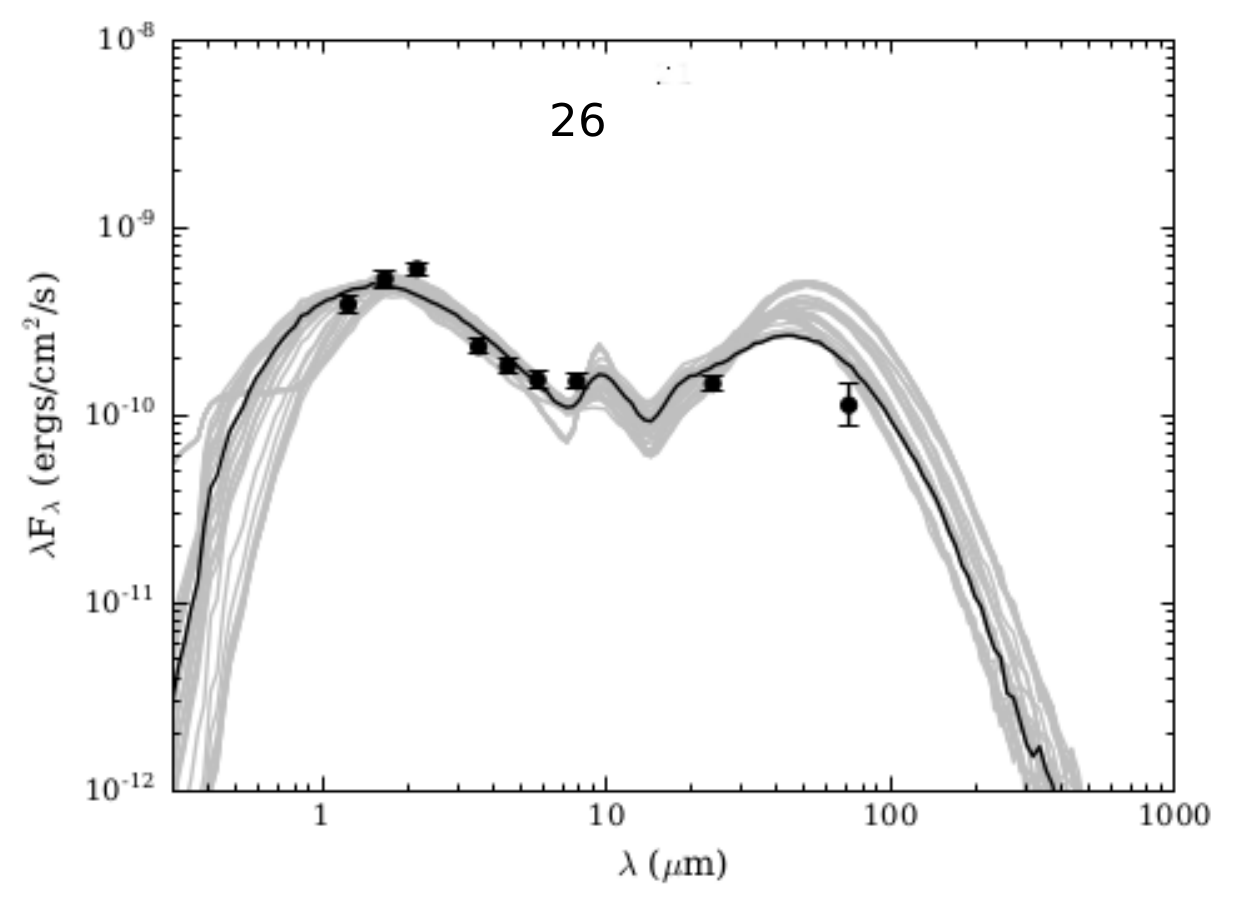}
\includegraphics[width=4.2cm] {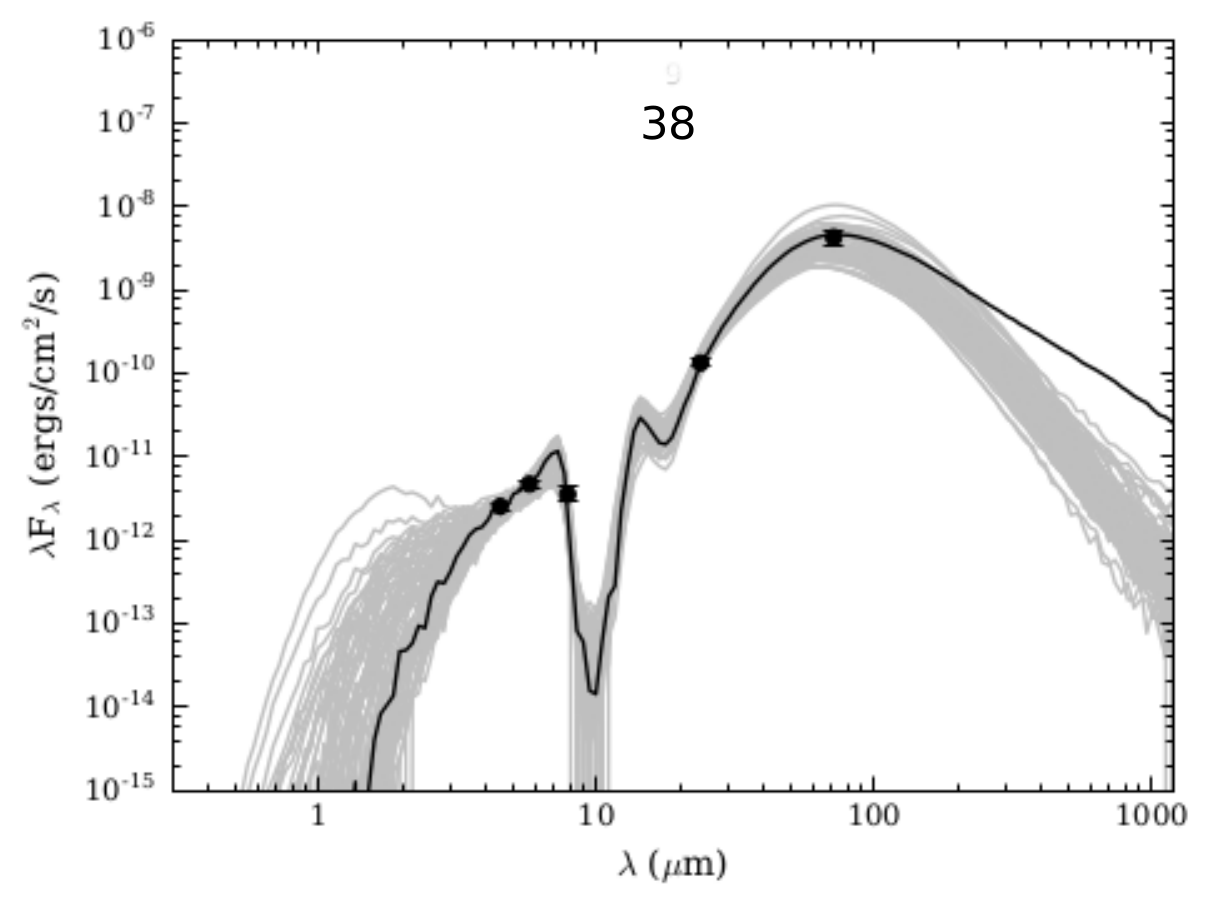}
\includegraphics[width=4.2cm] {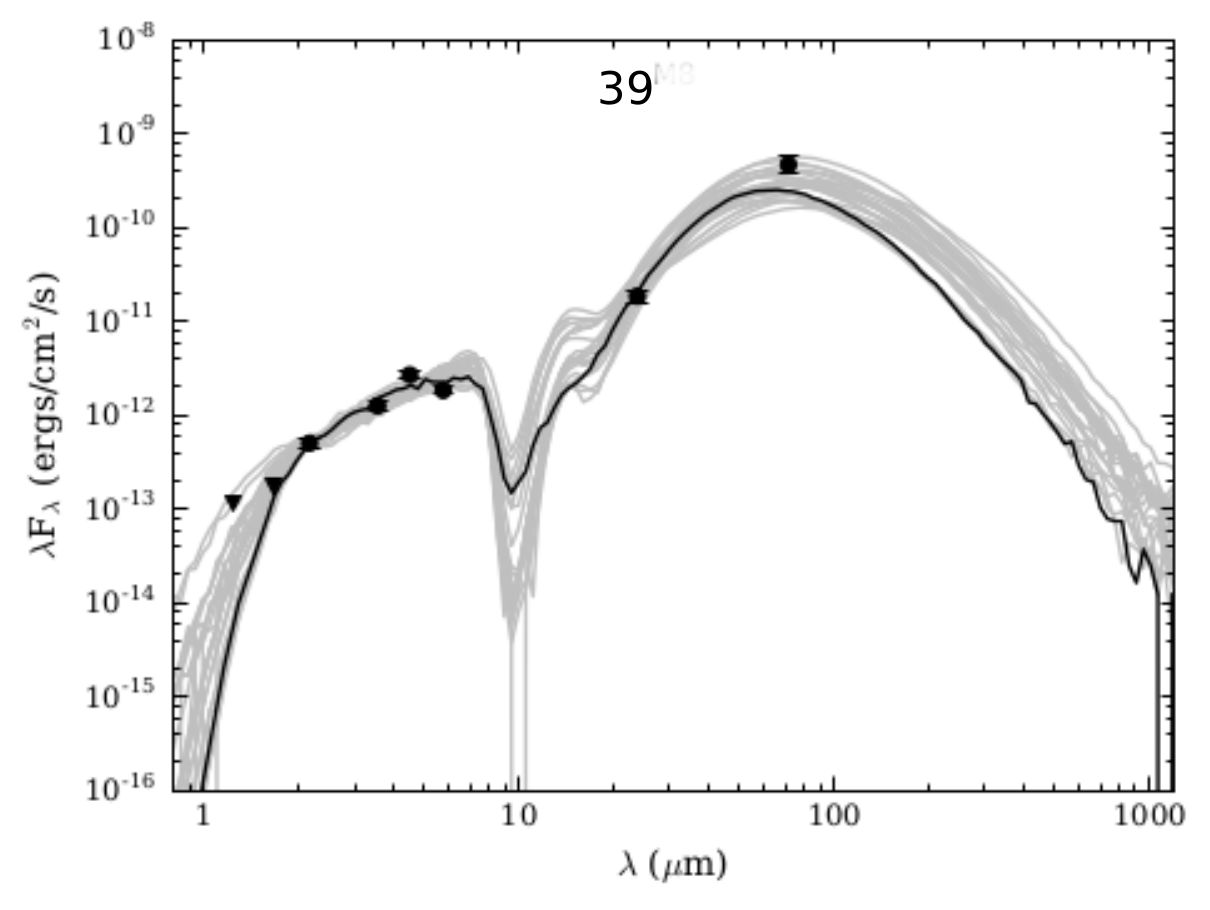}
\includegraphics[width=4.2cm] {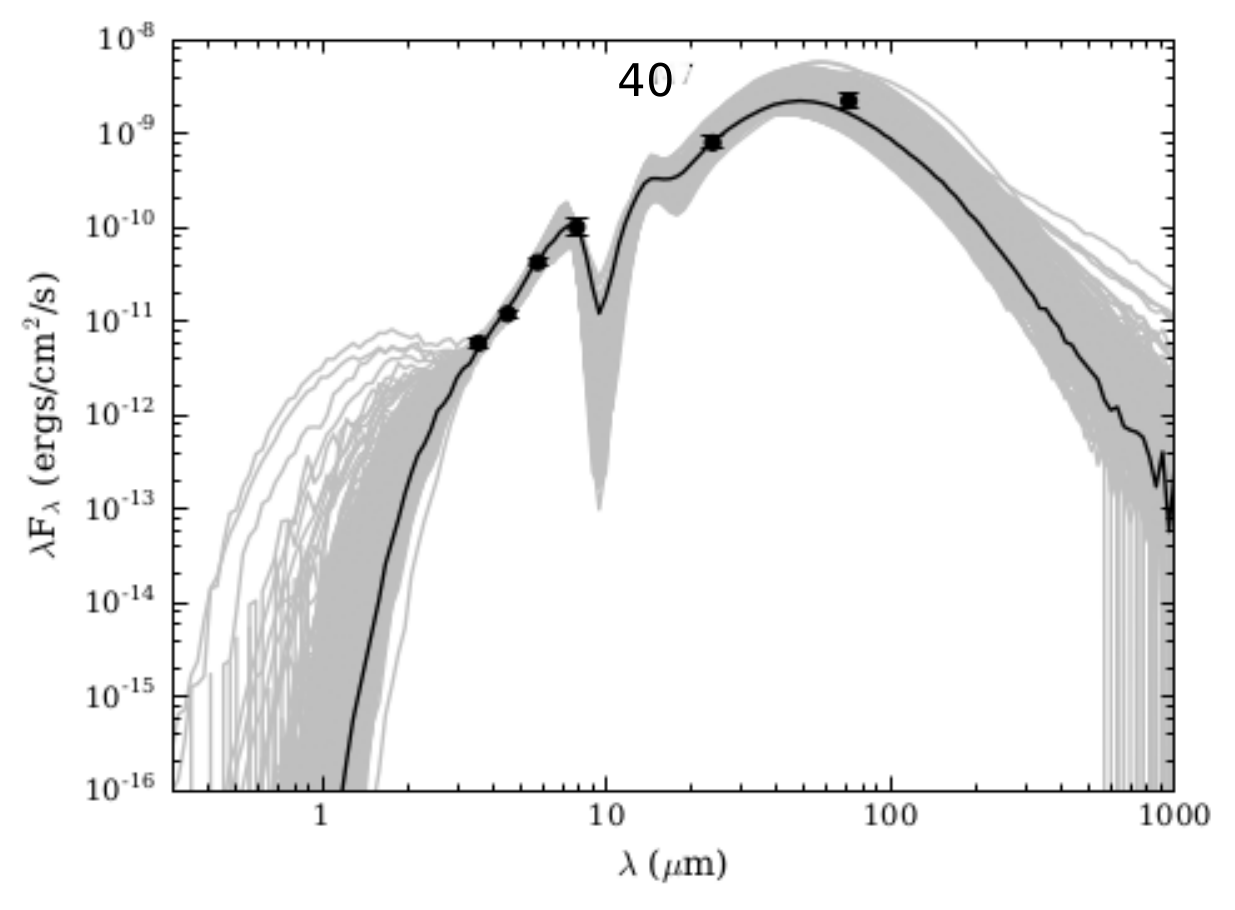}
\includegraphics[width=4.2cm] {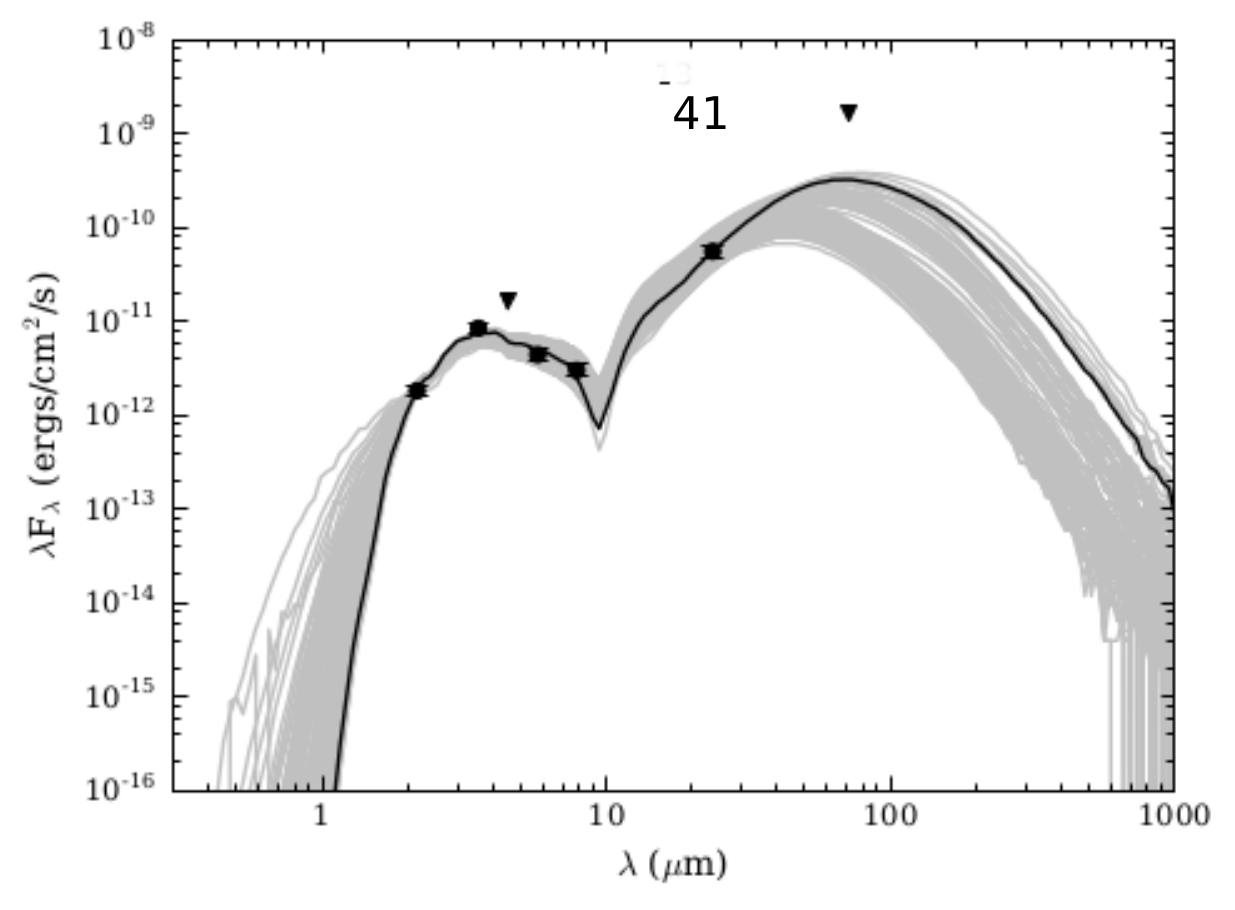}
\includegraphics[width=4.2cm] {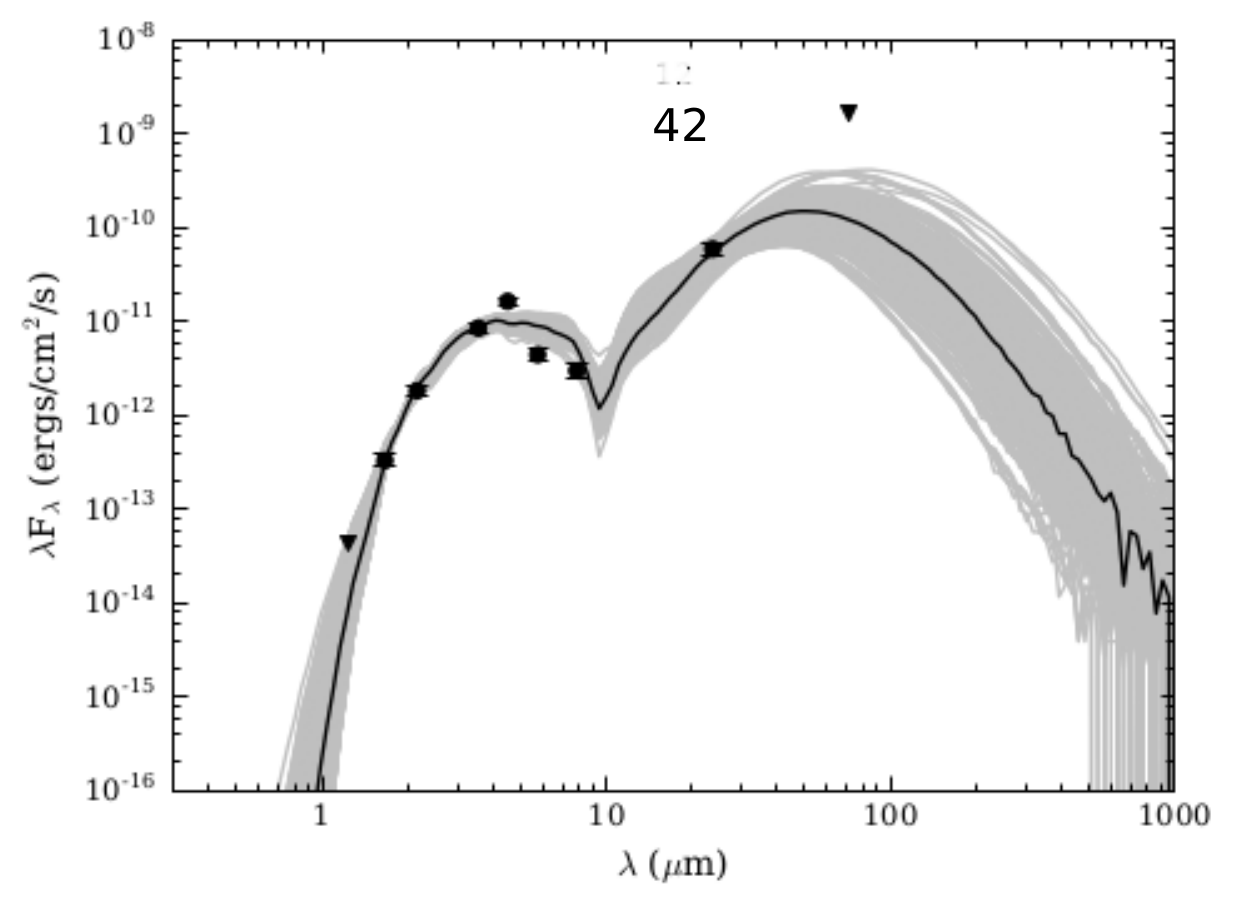}
\includegraphics[width=4.2cm] {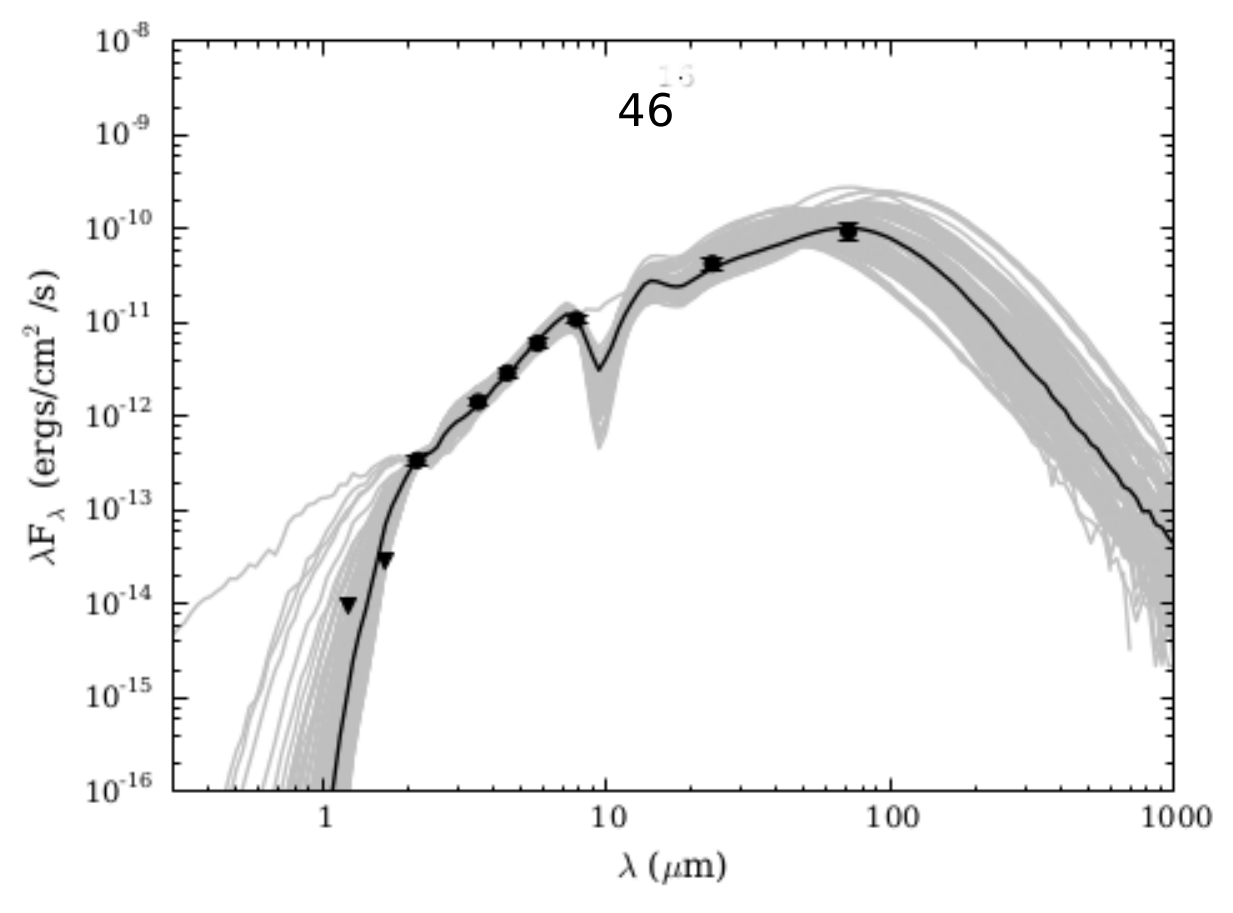}
\vspace{-0.0cm}
\caption{The observed SEDs of the driving sources and their best-fit models. The ID numbers correspond to the 
entries in the first column of Table. 1. The black line shows the best-fitting model, and the grey-lines show subsequent best-fitting models
that satisfy $\chi^2 - \chi^2_{\rm min} \leq 2N_{\rm data}$ criteria. 
Filled circles and triangles show the `data points' and `upper-limits' respectively. 
The IDs correspond to the MHO names given in Table. 1.}
\label{fig_seds}
\end{figure*}

For the SED fitting, we adopted a distance 2 $\pm$ 0.2 kpc  and 
a visual extinction (A$_V$) range 2 to 50 magnitude with lower-limit  corresponds  to 
the foreground extinction to M17 \citep[e.g.][]{hof08}, while the upper-limit in line 
with typical maximum values found towards EGOs \citep[e.g.][]{carati15} and \uchii~ regions \citep[e.g.][]{han02}. 
While fitting model SEDs, we adopted an error of 10, 15 and 20 per cent, respectively, for the UKDISS, {\it Spitzer} and 70 $\mu$m fluxes 
instead of formal photometric errors  in order to fit without possible bias that may cause 
by underestimation of the flux uncertainties.
Figure \ref{fig_seds} shows the model SEDs of 14 sources for which we have reasonable number of data points between 1.2 to 70 $\mu$m. As can be seen, although 
SED models show some degree of degeneracy, they appear to fit the data reasonably well. Barring source $26$ (i.e, 
the driving source of MHO 2333), all the SED models clearly show rising SEDs up to 70 $\mu$m, 
consistent with our earlier classifications that majority of them are protostars. 

Like any other models, these models have their own sets of limitations. For example, these models do not 
account for interstellar radiation fields (IRSF), cold 
dust of the protostellar envelopes (i.e. dust below 30 K) and  stellar multiplicity, etc. 
The main objective here is not to provide a precise set of physical parameters for the YSOs but to find the range of a few key 
parameters from the models and then 
discuss the possible nature of the sources. To do so, we obtained  physical parameters of the sources by
adopting the approach similar to \cite{rob07}, i.e. by considering those models  satisfy 
$\chi^2 - \chi^2_{\rm min} \leq 2N_{\rm data}$, where $\chi^2_{\rm min}$ is the 
goodness-of-fit parameter for the best-fitting model and $N_{\rm data}$ is the number of input observational
data points. We then obtained  the parameters from the weighted mean and standard deviation of these best-fitting  models weighted by
e$^{({{-\chi}^2}/2)}$ \cite[e.g.][]{sam12}. These parameters are tabulated in Table. 2, including  the stellar  mass (M$_{\ast}$),
stellar age (t$_{\ast}$), disc mass (M$_{\rm disc}$)
disc accretion rate ($\dot{M}_{\rm disc}$) and  total luminosity ($L_{\mathrm{bol}}$) of each source. 
As per the models, 
the disc masses are in the range 0.003 -- 0.14 \msun~ with a median  $\sim$ 0.02 \msun, 
disc accretion rates are in the range  0.08 -- 9.7 $\times$ 10$^{-7}$ \msun~ yr$^{-1}$ 
with a median $\sim$ 2.8  $\times$ 10$^{-7}$ \msun~  yr$^{-1}$, and age in the range 0.05 -- 3 $\times$ 10$^5$ yr, 
with a median value $\sim$ 1 $\times$ 10$^5$ yr. 

\begin{figure}
\centering
\includegraphics[width=8.5cm]{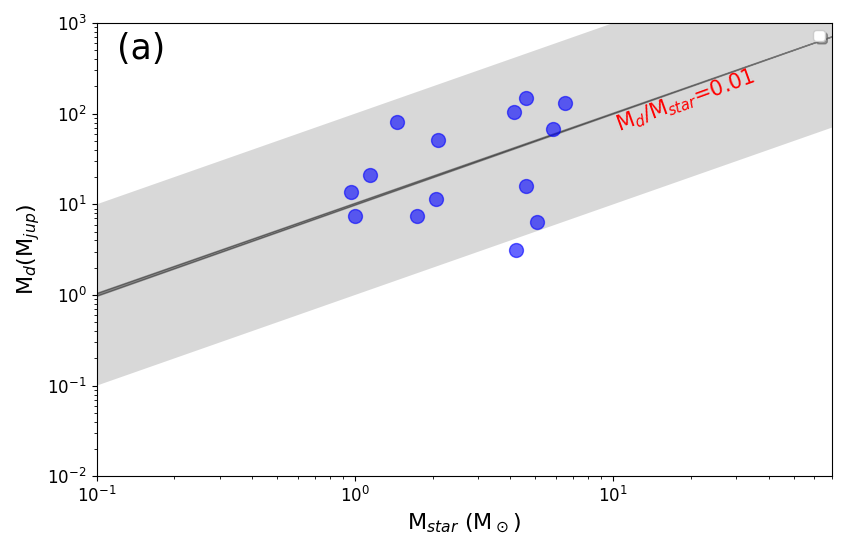}
\includegraphics[width=8.5 cm]{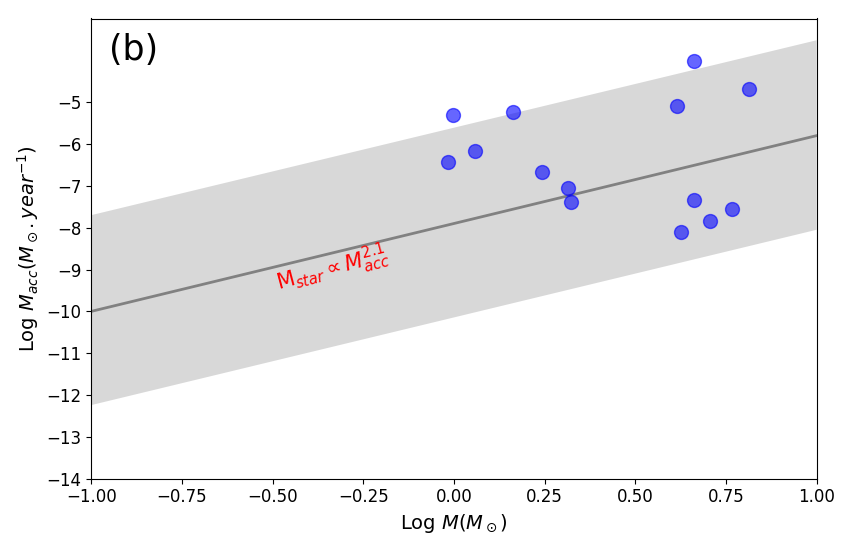}
\vspace{-0.2cm}
\caption{(a):  Variation of protoplanetary disc mass (in the unit of Jupiter mass) with the mass of the central
star. (b): Correlation between stellar-mass and accretion 
rate. In both plots, our SED model based measurements are shown in blue dots, 
while the  solid line and the shaded area represent the measurements 
from literature for the class II/III YSOs and
have the same meaning as described in the text.}
\label{sed_res}
\end{figure}

Figure \ref{sed_res}(a) shows the disc masses  of the YSOs 
obtained in the present work (blue dots) and their comparison with those in the nearby star-forming regions 
\citep[][]{will11}, measured through sub(mm) observations.
In Fig \ref{sed_res}(a), the solid 
line represents the median ratio of disc to stellar mass, i.e. $\sim$ 1 per cent as derived by \citet[][]{will11} 
from the compilation of a large number of class II/III YSOs 
in the stellar mass range 0.4 -- 10 \msun, and the shaded area represents the 1 dex spread about the median value of their sample. 
Similarly, Fig. \ref{sed_res}(b) shows the disc accretion rates of the YSOs obtained in 
the present work (blue dots) and their comparison with the literature values of the  
nearby star-forming regions as complied by \citet[][]{hart16}. These authors complied a large number of 
class II/III YSOs, for which  accretion rates are measured in various ways (e.g.  spectroscopic
measurements of the Balmer continuum, photometric $U$-band measurements and 
emission line measurements) and observed a strong correlation with stellar-mass.
In Fig. \ref{sed_res}(b), the solid line represents the relation,
$\dot{M}_{\rm disc}$  $\propto$ M$_{\ast}^{2.1}$,  obtained by \citet[][]{hart16}, and the 
shaded area represents the 3$\sigma$ scatter around this line, where $\sigma$ 
is $\sim$ 0.75 dex \citep[see][for detailed discussion]{hart16} .  

As can be seen from Fig. \ref{sed_res}, despite the evolutionary difference between the sources (i.e. protostars versus class II/III YSOs), 
in general, the SED model based disc measurements show a fair agreement with the literature values at any given stellar mass, 
indicating that the obtained disc properties of the driving sources may be the representation of their true values. In 
the absence of high-resolution  mm observations and precise extinction measurements, we emphasize that the 
obtained disc parameters are still indicative and should be treated with caution.  On the other hand, as one can see from  Fig. \ref{fig_seds},
for a given source the overall shape of all the models in general is similar, so bolometric luminosities of the sources are expected 
to be better constrained. In fact, we find they are in reasonable agreement with luminosity estimated based on  the 
70 $\mu$m flux alone (discussed in Section \ref{lyso}).

\begin{table*}
\scriptsize
\caption{\label{t:R07}Inferred Physical Parameters from the SED modeling}
\begin{tabular}{cccccc}
\hline\hline
 ID  &\multicolumn{1}{c}{$M_{\ast}$} & \multicolumn{1}{c}{age} & \multicolumn{1}{c}{$M_{\rm disc}$ }
& \multicolumn{1}{c}{$\dot{M}_{\rm disc}$ } & \multicolumn{1}{c}{${L}$}  \\

  &\multicolumn{1}{c}{($M_\odot$)} & \multicolumn{1}{c}{(10$^{5}$ yr)} & \multicolumn{1}{c}{( $M_\odot$)}
& \multicolumn{1}{c}{(10$^{-8}$ $M_\odot$/yr)} & \multicolumn{1}{c}{(10$^{2}$ $L_\odot$)}  \\
 \hline
\hline
01  & 2.10       $\pm$   0.87 &1.56      $\pm$   0.90 &0.049     $\pm$   0.073 & 00.04   $\pm$   00.09 & 0.27    $\pm$   0.06\\
06  & 1.75       $\pm$   0.70 &0.99      $\pm$   0.68 &0.007     $\pm$   0.009 & 00.21   $\pm$   00.64 & 0.26    $\pm$   0.09\\
10  & 4.13       $\pm$   0.70 &0.14      $\pm$   0.26 &0.099     $\pm$   0.057 & 08.24   $\pm$   06.46 & 2.52    $\pm$   0.17\\
15  & 1.45       $\pm$   0.62 &0.15      $\pm$   0.31 &0.077     $\pm$   0.037 & 05.78   $\pm$   06.30 & 0.53    $\pm$   0.22\\
19  & 2.07       $\pm$   0.53 &1.32      $\pm$   0.64 &0.011     $\pm$   0.013 & 00.09   $\pm$   00.20 & 0.34    $\pm$   0.09\\
23  & 0.96       $\pm$   0.31 &0.86      $\pm$   0.69 &0.013     $\pm$   0.013 & 00.36   $\pm$   00.97 & 0.13    $\pm$   0.04\\
24  & 1.00       $\pm$   0.32 &0.05      $\pm$   0.28 &0.007     $\pm$   0.012 & 05.04   $\pm$   09.42 & 0.35    $\pm$   0.13\\
26  & 5.85       $\pm$   0.17 &2.77      $\pm$   0.77 &0.065     $\pm$   0.042 & 00.03   $\pm$   00.03 & 2.86    $\pm$   0.44\\
38  & 4.58        $\pm$   2.50 &0.04      $\pm$   0.07 &0.141     $\pm$   0.107 & 97.49   $\pm$   95.73 & 7.15    $\pm$   1.57\\
39  & 5.07        $\pm$   0.61 &2.10      $\pm$   0.63 &0.006     $\pm$   0.016 & 00.01   $\pm$   00.03 & 1.47    $\pm$   0.43\\
40  & 6.51        $\pm$   1.75 &0.78      $\pm$   0.71 &0.126     $\pm$   0.146 & 20.39   $\pm$   40.12 & 8.27    $\pm$   2.06\\
41  & 4.60       $\pm$   0.26 &1.40      $\pm$   0.08 &0.015     $\pm$   0.010 & 00.05   $\pm$   00.01 & 0.92    $\pm$   0.06\\
42  & 4.22       $\pm$   0.79 &2.76      $\pm$   1.10 &0.003     $\pm$   0.007 & 00.01   $\pm$   00.04 & 0.70    $\pm$   0.14\\
46  & 1.15       $\pm$   0.31 &0.25      $\pm$   0.17 &0.020     $\pm$   0.017 & 00.67   $\pm$   01.52 & 0.20    $\pm$   0.06\\
\hline
\end{tabular}
\end{table*}

\subsubsection{Luminosity of the driving sources using only 70 $\mu$m flux} \label{lyso}

\citet{dunham06} using radiative transfer models demonstrated that 70 $\mu$m is a crucial wavelength for determining  $L_{\mathrm{bol}}$ for 
embedded protostars, as radiative transfer models are  strongly constrained by 
this wavelength, and  it is largely unaffected by the 
details of the source geometry and external heating. Furthermore, 
\citet{dunham08} examined c2d protostars and obtained the following tight correlation between $L_{\mathrm{bol}}$ 
(excluding luminosity arising from external heating)  and  F$_{70}$ for low-luminosity protostars:
\begin{equation}
L_{\mathrm{bol}} = 3.3 \times 10^8 \, F_{70} \, ^{0.94} \,(d/140 pc)^{2}\, \lsun \quad,
\end{equation}
where $F_{70}$ is in cgs units (erg cm$^{-2}$ s$^{-1}$). 
While this method
is accurate within a factor of 2 to 3 \citep[\eg][]{dunham08,comm12}, it offers a proxy way of obtaining $L_{\mathrm{bol}}$ for 
embedded protostars.  It is worth noting that,
\cite{raga12},
have shown that the 70 $\mu$m flux of more massive sources also correlates well with their total luminosity and
there is a good agreement between the correlations established for low and high-luminosity sources. 
Among our YSO sample, five sources have detection only in 24  and 70 $\mu$m, or in 70 $\mu$m, 
on which we could not perform SED modeling. We therefore used the above empirical relationship
for obtaining bolometric luminosity of these  candidates. 
Of these, four are found to be in the range of 
$\sim$ 4--53 \lsun, and one is exceptionally luminous with  luminosity $\sim$ 0.5 $\times$ 10$^4$ \lsun. The latter source 
corresponds to an extremely young YSO associated to an EGO (see Section \ref{notes_mho} for further discussion). The luminosity 
of these sources is also tabulated in Table. 2. We  compared the 70 $\mu$m flux based and SED model based luminosities for 
common protostars (i.e. class 0/I), and found that two estimates are in agreement with each other within a factor of 2. 

From the luminosity and the disc properties of the driving sources, we can say that most of them are intermediate-mass 
young stellar objects (for mass-luminosity relationship of protostars, see Fig. 4 of \citet[][]{and08} and references therein) that are still actively 
accreting from their protoplanetary disc. We note, outflow power is stongly
correlated to YSO luminosity \citep[e.g.][]{carati15}, therefore the predominance of moderate-mass YSOs in our sample of driving sources could be 
a selection effect caused by  sensitivity of the {\it Spitzer} and {\it Herschel} images at the distance of M17.

\section{Discussion}
\subsection{Notes on general nature of the MHOs and discussion on a few types of interesting sources} \label{notes_mho}
In this work, we have identified 48 MHOs. Of which, 45 (93 per cent) MHOs are new discoveries. 
Of the three already known outflows, two (MHO 2306 and MHO 2344) have been identified by \citet{cygan08} as EGOs using {\it Spitzer} observations 
and one  (i.e. MHO 2307) by \citet{carati15} using 2.12 $\mu$m observations. This shows the 
improvement UWISH2 can bring when it comes to identify outflows over galactic scales. 

Out of 48 MHOs, we could only associate 20 YSO candidates. This corresponds to 40 per cent of all the MHOs. If we consider the dust cores and clusters as the 
potential driving sources, the number of sources increases to 30 which is $\sim$ 60 per cent of all the MHOs.
We could not link the remaining MHOs with any potential YSO candidates, which could either be due to indistinct 
shock orientation with respect to the nearby YSOs or shocks are from distant sources or 
shocks are from low-luminosity sources beyond our sensitivity limit ($<$ 3 \lsun; discussed below). But largely the situation 
here is similar to  Serpens and Aquila \citep{ioa12a}, Orion A \citep{davis09}, Vela C \citep{zha14} and Cygnus-X \citep{mak18}, 
where only 50 to 60 per cent of the MHOs were found to have associated YSOs. 

From SED models or using only 70 $\mu$m fluxes, we find that the outflows are mostly driven by sources of  
luminosity in the range 4--1000 \lsun, suggesting that more low-to-intermediate mass YSOs with outflows can be studied and 
characterized over larger distances with the help of UWISH2 survey. 

In our sample, five (MHO 2306, MHO 2328, MHO 2329, MHO 2331, MHO 2337 and MHO 2343) out of the 20 YSOs have no detection in 
5.8 and/or 8.0 $\mu$m bands in the MIPSGAL catalogue, yet were detected in 24 and/or 70 $\mu$m.
Our findings suggest that search for outflow driving source without 24 and 70 $\mu$m 
would result in missing  many potential candidates (e.g. $\sim$ 30 per cent in the present case). 
24 $\mu$m is an important wavelength for determining whether a source features 
a rising (or falling) SED. 70 $\mu$m is particularly important for the identification of very young protostars and  
perhaps even for first hydro static cores \cite[e.g.][]{enoc10}. We find as protostars are bright
at mid- and far-infrared, in the multi-colour IRAC-MIPS images the outflow-driving sources clearly stand out compared to 
the nearby sources (e.g. see Figures in Appendix B).

\begin{figure}
\includegraphics[width=\columnwidth]{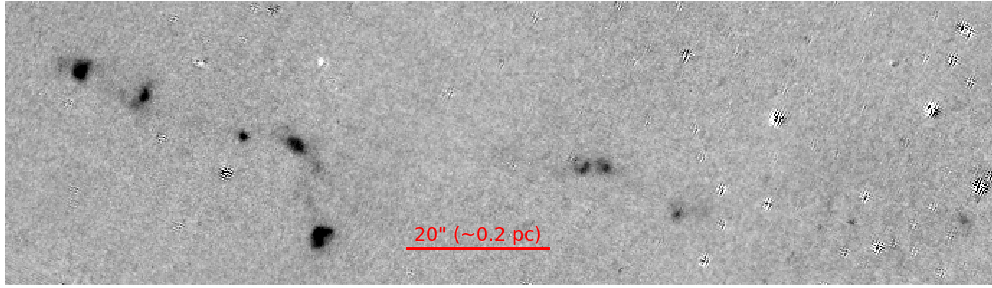}
\vspace{-0.5cm}
\caption{Continuum subtracted \htt image of the MHO 2347, showing chain of knots spread over parsec-scale. We note, the bright bow-shock shaped knot
seen at the central bottom part of the figure is part of MHO 2307. 
}
\label{fig_var}
\end{figure}

In our sample (see Table. 1), although the majority of the bipolar outflows have a total length less than 0.4 pc, the MHO  2347, however, 
corresponds to a parsec-scale outflow (see Fig. \ref{fig_var}) from a class 0 source. We note,
\citet{carati15} have also observed  MHO 2347, but only part of it. With
our sensitive observations we detect additional three knots along the southwestern lobe of MHO 2347, which makes it 
the largest \htt outflow of the M17 complex. 
We find the outflow shows some degree of bend morphology and
consists of a series of compact knots at a median spacing of $\sim$ 0.15 pc (for details see Appendix B). 
Episodic ejection of material by quasi-periodic outbursts of the central star-disc system 
is one of the possible explanation of regular spacing between the knots 
\citep[e.g. ][]{raga10}. Recently, \citet{vor15} using hydrodynamical simulations showed that
episodic accretion events induced by gravitational instabilities and disc fragmentation, 
are present mainly during the early evolution (class 0) of most protostellar systems. 
In addition to bursts induced by disc instabilities, it has also been suggested that bursts can 
be induced by external interactions with one or more companion stars. So objects like MHO 2347
are of particular interest for studying the accretion history and cause of variability in 
very young systems \cite[e.g. see discussion in][]{mey17,her17,car17}.

\begin{figure}
\centering
\includegraphics[width=7cm]{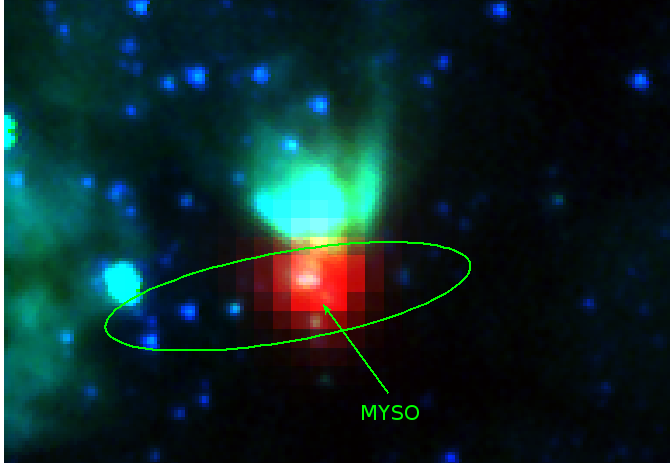}
\vspace{-0.0cm}
\caption{Color composite image around  MHO 2306 at 3.6 (blue), 8.0 (green) and 70 $\mu$m (red). 
The ellipse denotes the flow axis of the MHO 2306 (for details see Appendix B). The 
arrow points to the 70 $\mu$m bright MYSO, the likely source responsible for MHO 2306.}
\label{fig_myso}
\end{figure}

In our list, we find that MHO 2306 corresponds to an EGO (details given in  Appendix B). This MHO has been identified by \cite{lee12} using UWISH2 images while searching 
for EGOs counterpart in \htt. We find a bright 70 $\mu$m point source at location of EGO. The source location also coincides with the location 
of a methanol maser \citep{bay12} which is a tracer of early evolutionary stages of high-mass star formation \citep{ell06}.  The source lies 
adjacent to a compact \hii region that is bright at 8 and 24 $\mu$m (see Fig. \ref{fig_myso}) and  has no  24 $\mu$m counterpart 
in the MIPSGAL catalogue. It is embedded in a clump of column density $\sim$ 7$\times$ 10$^{23}$ \cms \citep{tang17}, thus
in a region of very high extinction. Its 70 $\mu$m flux suggests to a source of luminosity $\sim$ 0.5$\times$10$^4$ \lsun,
however, no associated radio free-free emission at 5 GHz (resolution $\sim$ 1.5 arcsec and rms $\sim$ 0.4 mJy) is found 
in the CORNISH\footnote{http://cornish.leeds.ac.uk/public/index.php} survey image \citep{hora12}.
It appears that, although its luminosity represents a high-mass star, 
it is yet to develop an \uchii~ region. Thus represents a young massive YSO with an outflow.
Similarly, MHO 2344 (driven by a $\sim$ 700 \lsun~ class 0 type YSO) present a case where mutiple 
wide-angle bullets have been observed (see Fig. \label{fig_mh009} and Section \ref{iden_mho} for details). We suggest, these massive YSOs 
are potential candidates for studying  various aspects of early phases of massive star-formation and evolution 
\cite[e.g. see discussion in][]{tan16,bal17}.

\begin{figure*}
\includegraphics[width=11cm]{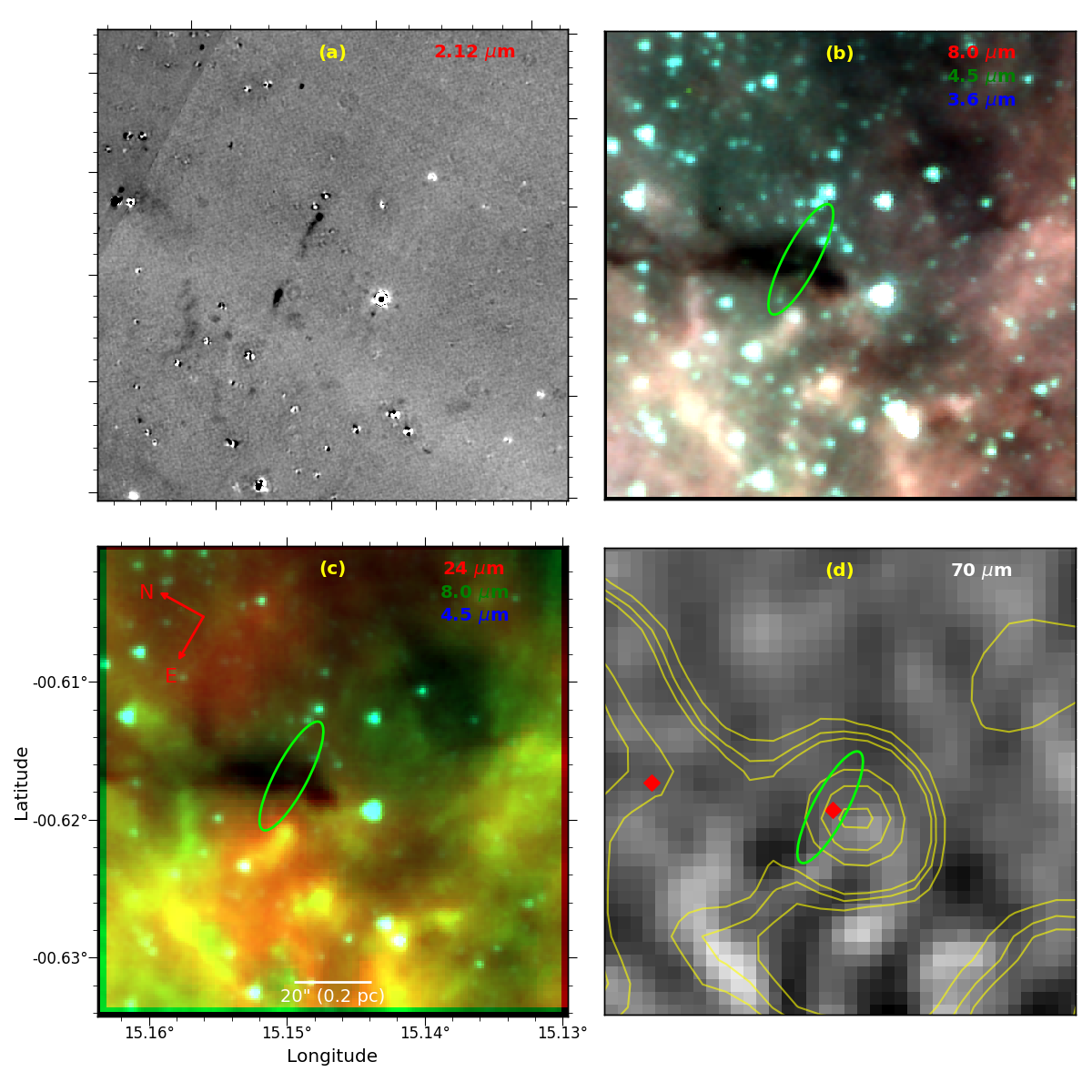}
\vspace{-0.5cm}
\caption{Multicolour image around the MHO 2352, showing the emergence of \htt jets from a {\it Spitzer} dark-cloud, with no
embedded point sources up to 70 $\mu$m. The colour codings have the same meaning as in Fig. \ref{fig_mho23}. The diamond symbols
on the 70 $\mu$m image represent the SCUBA cores. The 870 $\mu$m contours are shown in orange colours. 
}
\label{fig_mho03}
\end{figure*}

We identified \htt jets and knots in the vicinity of six 70 $\mu$m dark cores (MHO 2335, MHO 2338, MHO 2341, MHO 2350, MHO 2352 and MHO 2353),
but for three  cases (MHO 2341, MHO 2350 and MHO 2352), we observed  symmetric jets clearly emanating 
from the cores located at the middle of the jet axis. 
Figure \ref{fig_mho03} illustrates an example. As can be seen, the MHO consists of two strong opposite jets. 
Together, they delineate a north-south outflow. In the {\it Spitzer} bands, a dark-cloud perpendicular 
to the flow axis, can be seen as an absorption feature against the bright background.  
The morphology  and the compactness of the jets strongly suggest that a driving source should be 
embedded inside the dark-cloud. No point sources were found between 3.6 to 70 $\mu$m in the dark-cloud along the axis of the jets. 
However,  one can see a SCUBA core (mass $\sim$ 9 \msun; \citet{reid06}) lies at the expected location. Thus, the core is the most 
likely source responsible for the jets. Most of the driving sources identified in this work are bright 
in mid-IR and visible in 70 $\mu$m. The 3$\sigma$ point-source sensitivity of the 70 $\mu$m image is $\sim$ 0.24 Jy, where $\sigma$ is the  
standard deviation of the background intensity. Using equation (1),  0.24 Jy corresponds to a luminosity $\sim$ 3 \lsun. 
Thus, we cannot ignore the possibility that the starless cores may in fact harbor faint YSOs of  luminosity $<$ 3 \lsun. 
At this point, we are  ill-qualified to comment whether these 70 $\mu$m dark cores are starless or protostellar. 
None the less, these mid- and far-infrared quiet clumps/cores are  potential targets for the 
understanding of early phases of core collapse and fragmentation \cite[e.g. see discussion in][]{tra18,pal18}.

\subsection{Dominant YSO class responsible for the jets} \label{dom_yso}
The present \htt  survey cover a small area about 1.5 square degree, but it is worthwhile 
to compare our results with the results from other similar surveys of  nearby star-forming regions
to understand at what stage of protostellar evolution \htt jets are prominent.
In the present work, we find that of the 20 YSOs with outflows, 18 ($\sim $90 per cent) outflows are driven by protostars 
(i.e, class 0/I YSOs), which is similar to the $\geq$ 80 per cent of the  \htt outflows in Orion A \citep{davis09},
$\sim$ 90 per cent in
Corona Australis \citep{kumar11}, $\geq$ 90 per cent in Serpens South \citep{teix12}, 
$\sim$ 70 per cent in Aquila \citep{zhang15} and $\sim$ 80 per cent in Cygnux-X \citep{mak18} molecular clouds,  
except the Cassiopeia and Auriga complex \citep{fro16} where 20 per cent of the driving sources are protostars, the remainder 
are Classical T-Tauri Stars. 
The above trend (except Cassiopeia and Auriga complex) hints that the \htt jets are mostly prominent in class 0/I type protostars and the jet 
activity possibly decreases significantly as the object evolves. However, we stress that the above trend can be biased by the 
selection effect as older sources tend to reside in relatively molecule-free environments, and thus their shocks may not be 
traced by MHOs. Sensitive optical observations at H$_\alpha$ and [SII] lines would be helpful to solidfy the above trend.

\subsection {Jet lengths and  dynamical time scales } \label{jet_length}
\begin{figure}
\centering
\includegraphics[width=8.5cm]{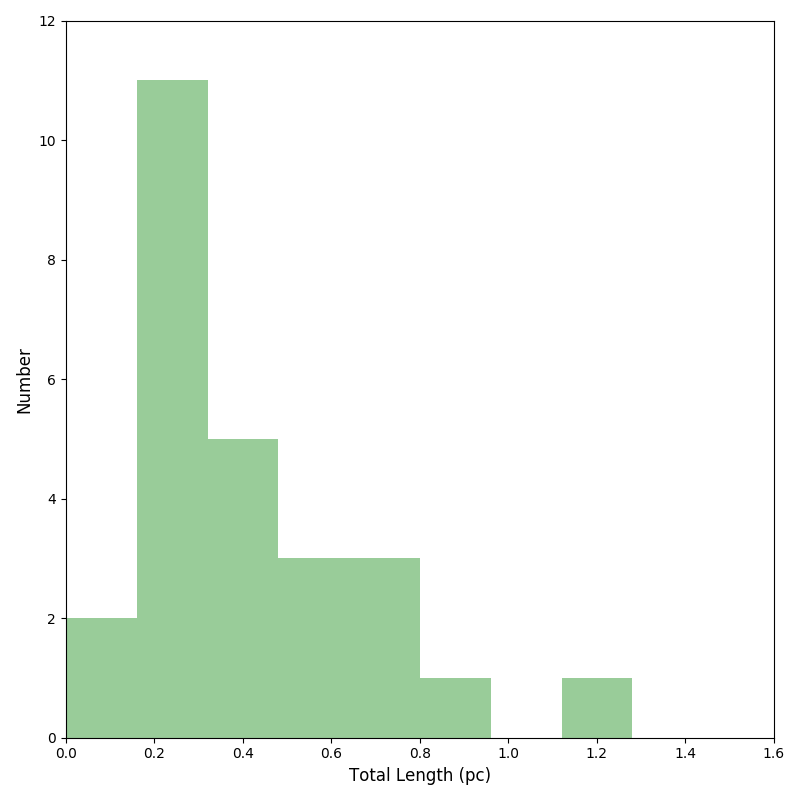}
\vspace{-0.3cm}
\caption{Total length distribution of the  outflows that are either associated with a YSO or a core.}
\label{fig_hist}
\end{figure}

Figure \ref{fig_hist} shows the frequency distribution of total length of the outflows that are associated either with a YSO or a core. 
Here, we assume that the total length of a monopolar appearance outflow is
twice of its single-sided lobe length (the distance of the farthest \htt feature from 
the YSO/core), while it is end-to-end for bipolar outflows. Using this simple approach, we find that the total 
outflow lengths of the MHOs are in the range of 0.05 pc to 1.2 pc, with a median  $\sim$ 0.34 pc. 

We note that these lengths refer to the lengths of the visible  \htt jets, 
not necessarily the total lengths of the flows, as  
outflow length depends on many factors such as the tracer in question, inclination angle to the line of sight,
the density of the ambient medium, etc., so more extreme values may be  feasible. 
\citet[][]{fro16} argued that, since bipolar outflows are often asymmetric, 
measuring outflow lengths in the traditional way (from end-to-end) should be avoided. 
Owing to low statistics of the outflow sources, here we did not measure the length of each lobe separately. None the less,
if we consider the length of the longest lobe as the true half length of a bipolar flow, then we find the total outflow
lengths of all the MHOs are in the range of  0.09 pc to 1.6 pc, with a median  $\sim$ 0.44 pc. 
We would like to point that outflows can break-out of their parent clumps and interact with medium of little molecular
gas; thus, the MHO-based outflow sizes can be limited by the spatial extent of
the molecular cloud, hence are likely representative of lower-bounds of the true extent of the flows.
We estimated the likely  lower-limits of the dynamical time-scales ($t_{dyn}$) of the outflows using the relation  $t_{dyn}$ = D/($V$ sin i), 
where D is the length of the lobe, V is the velocity  of the jets, 
and i is the inclination of the outflow with respect to the line of sight.
Here we assume an average flow inclination angle  $\sim$ 57$^\circ$.3 \citep[see discussion in][]{bon96},  
a flow velocity  $\sim$ 35 \kms~ which is the median value of  86 \htt features measured by 
\citet{zhang13} using proper motion observations, and jets travel with a constant 
velocity since their launch. Under these assumptions, we find that the time-scales of ejection are  
in the range 0.01--0.2 $\times$ 10$^5$ yr, with a median $\sim$ 0.07 $\times$ 10$^5$yr. The median time-scale 
is slightly lower than the lifetime of the class 0/I sources ($\sim$ 0.4 $\times$ 10$^5$yr; \citealt{evans09}) and the median age of
the YSOs as estimated from the SED modeling. Proper velocity and size measurements, and inclination 
angle of the jets would shed more light on the correlation between the dynamical time-scale and lifetime of the protostars.

\section{General picture of star formation in the M17 complex} \label{over_sf}
As discussed in Section \ref{dom_yso}, jets are possibly
more prominent in the class 0/I phase of a YSO. Thus the distribution of \htt jets/knots offers an independent 
diagnostic to trace the recent star formation activity of a complex. Fig. \ref{fig_ovpic}a shows relative spatial distribution 
of the MHOs and the cold gas at 870 $\mu$m. As can be seen, the cold dust emission (red contours) 
mainly corresponds to the absorption features of the background
24 $\mu$m image, and primarily concentrated at the location of A, B, C and D components of the complex. 
Also most of the MHOs are located in the close vicinity of the intense 870 $\mu$m emission, and  
they show a good correlation with the distribution of SiO emission (see Fig. \ref{fig_ovpic}b). 
We observed that in 70 per cent of the cases, the clump with a SiO emission coincides 
with an MHO, suggesting active star-formation is ongoing in  these  clumps. 
Looking at the large-scale spatial distribution of MHOs and 870 $\mu$m dust emission, it appears that 
M17 complex consists of several scattered star-forming clumps with jets and SiO emission, 
separated by distances of several parsecs from  each other. This suggests the hierarchical nature 
of the molecular cloud \cite[e.g. see discussion in][]{vaz17,cal18}.

One can also notice that among the structures of M17, M17SWex is  filamentary 
and \htt flows are more abundant in its direction, compared to other parts of the complex. 
Using high-resolution \amo~ and 1.3 millimeter 
observations, dense filamentary clouds with  density as high as $\sim$ 10$^{23}$ \cms 
have been observed in M17SWex \citep{bus13,bus16}. This high column of matter could have 
limited our ability to detect some of the weaker flows compared to the less extincted regions of the complex. 
Even so, based on the distribution of the MHOs, largely it appears that M17SWex is currently 
forming stars actively among all the fragments, thereby supporting the fact that in molecular clouds
filaments and filamentary structures are prime sites of active star formation \citep[e.g.][]{kon15,sam15,ray18}.
We find our results are  in accordance with the  results obtained  by \citet{pov10,pov16}, \citet{bus13} and 
\citet{bus16} for M17SWex. These authors using multiwavelength observations found that the M17SWex region is 
associated with a rich number of YSOs/cores, and suggest that it may form 
many more stars by accreting a significant amount of surrounding gas fed by the filaments.

In contrast to M17SWex, in  the interior of the \hii region, 
we observed only a few MHOs -- which is not surprising, given the fact that it hosts a 3 Myr old cluster 
(cluster centre is marked with a star symbol in the figure), 
where one would expect a low-fraction of sources with active 
accretion disc \cite[e.g.][]{jos17}. This implies that star formation has stopped or is in a more advanced phase
in the central region of the M17. The presence of a significant amount of  870 $\mu$m cold dust 
and a small number of outflows (either MHOs or SiO emission) surrounding the \hii region 
(particularly in the fragment 'B'), however, suggests that a very early phase of star formation 
is still ongoing at the periphery of the \hii region. 
We suggest that these outflows are probably from the second generation young 
stars whose formation has been triggered by the compression of the expanding \hii region, as observed
at the borders of several Galactic \hii regions \citep[e.g.][]{get12,pan14,samal14,deh15, ber16}. In fact,
strong evidence of M17SW and nearby molecular clouds are heated and compressed by 
the \hii region has been observed using molecular observations \citep[e.g.][]{lada76,wil03}.

All in all, in the M17 complex, M17SWex, appears to be the most active site of star formation,  
representing an excellent cloud for understanding  the early phases of cluster 
formation and evolution under the influence of strong outflow feedback.

\section{Summary}
In this work, we use the infrared observations of the \htt 1-0 S(1) 
line at 2.12 $\mu$m, with the UKIDSS, {\it Spitzer} and {\it Herschel} maps in 
the wavelength range 1.2--70 $\mu$m to identify protostars with jets and knots in
the M17 complex as well as to understand its ongoing star formation activity.

We identified 48 MHOs over 2.0\degree $\times$ 0.8\degree~area 
of the complex, i.e. potential  outflow candidates, $\sim$ 93 per cent of which are new
discoveries.  Based on the alignments and morphologies of the MHOs 
with the nearby YSO candidates, we could associate 20 YSO candidates to the outflows. 
Using  various flux ratios, we  deduce an evolutionary status of 
these candidates and find that  $\sim$ 90 per cent of the driving 
sources are protostars (\ie class 0/I), and 
only $\sim$ 10 per cent of the MHOs are driven by class II YSOs. Among the protostar, three
are likely PBRs.

Using the grid models of \cite{rob06}, we matched the model spectral 
energy distributions to the observed SEDs of the 14 outflow driving YSOs. This allows 
for the estimation of the physical properties such as  mass, luminosity, 
and accretion rate of the protostars. 
We find that the disc masses and disc accretion rates of $\sim$ 80 
per cent YSOs are in the range 0.003--0.14 \msun~ and 0.08--9.7 
$\times$ 10$^{-7}$ \msun~ yr$^{-1}$, respectively.  From SED modeling and using only 70 $\mu$m flux,
we estimated that the outflows are mostly driven by sources of 
luminosity in the range 4--1000 \lsun, suggesting that more low-to-intermediate mass YSOs 
with outflows can be studied and characterized over larger distances with the 
help of UWISH2 survey.

Our results show that six outflows are possibly emanating from  cores, where, 
no infrared sources were detected up to 70 $\mu$m. These sources are  important targets 
for follow-up studies for the understanding of the very early phase of star-formation. 

We observed a parsec scale bipolar outflow from a class 0 YSO with regularly spaced knots, a
potential candidate for understanding the variability in very young systems. We also observed
a strong spatial correlation between \htt jets/knots and SiO emission for massive ATLASGAL clumps.

Among the structures of M17, we find that  \htt jets/knots are 
statistically more numerous in the M17SWex region of the complex. Since \htt emission is
a tracer of recent ejecta from young protostars, we suggest that in the M17 complex, 
currently M17SWex is the most active region of star formation and is an excellent template
for the understanding of the early evolution of young clusters that are under the influence 
of strong outflow feedback.


\section*{Acknowledgments}
We are grateful to the anonymous referee for useful
comments that have helped us to improve the scientific
contents of the paper. 
M.R. Samal acknowledges the Ministry of Science and Technology of Taiwan for his post-doctoral fellowship. 
The data reported here were obtained as part of the UKIRT Service Program.


\appendix
\section{Details of the Classification of Young Stellar Objects} \label{iden_yso}
\begin{figure}
\centering
\includegraphics[height=6.5cm]{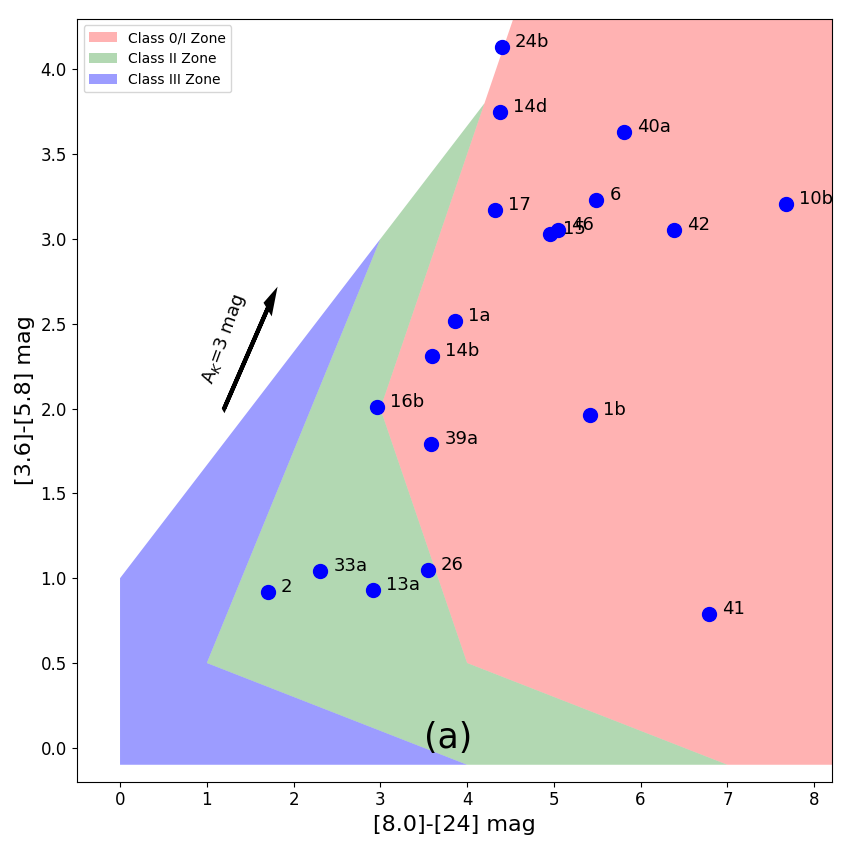}
\includegraphics[height=6.5 cm]{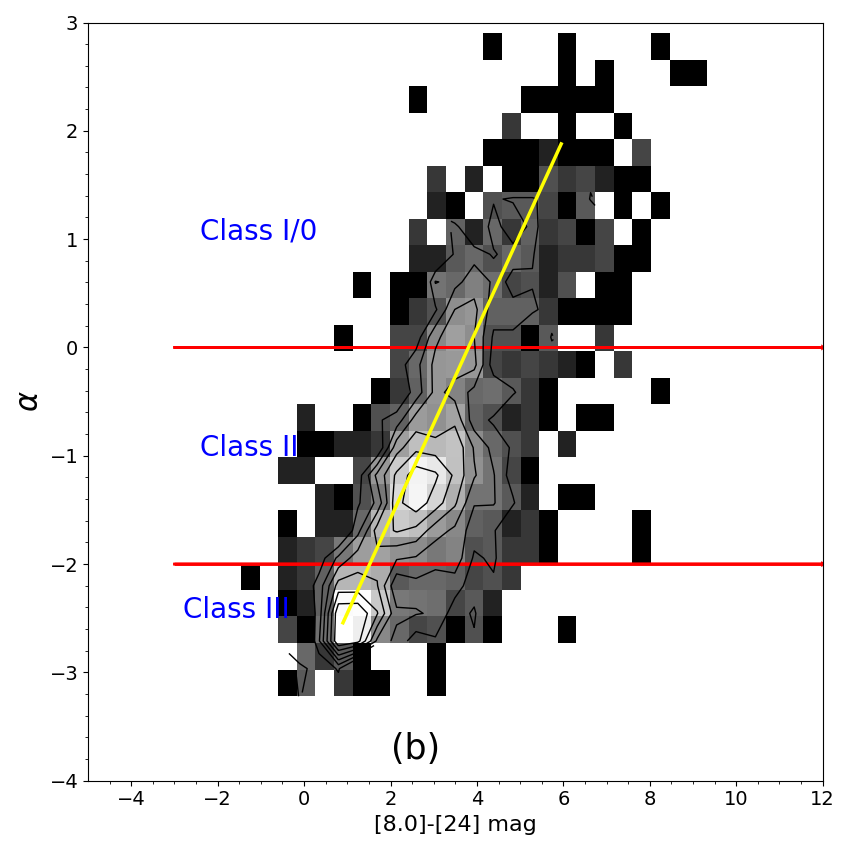}
\includegraphics[height=6.5 cm]{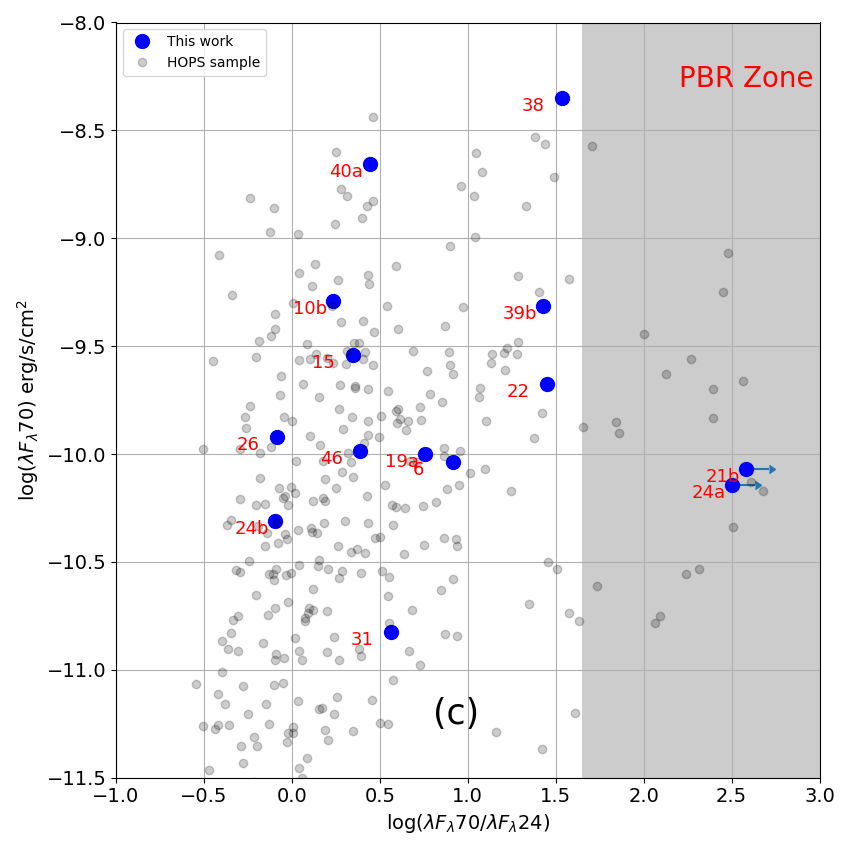}
\caption{
Classification of the potential driving sources with different criteria. 
(a): [3.6] -- [5.8] versus [8.0] -- [24] colour-colour plot of all the potential  
outflow driving sources (blue dots). (b): Hess diagram between [8] -- [24] colour and $\alpha$ of the \protect\cite{dun15}'s YSO sample. 
The yellow line corresponds to the linear fit to the peak of [8] -- [24] colour distribution, binned in  $\alpha$.  
(c): 70 $\mu$m flux versus  70 $\mu$m to 24 $\mu$m flux ratio for the potential
outflow driving candidates (blue dots). Sources without 24 $\mu$m detection 
are indicated with arrow marks. 
The shaded area (i.e. $log (\lambda F_\lambda 70)/(\lambda F_\lambda 24)$ $>$ 1.65) represents the zone where 
early class 0 objects (or PBRs) lie. The protostars from the HOPS sample are shown in grey dots.}
\label{iraccc}
\end{figure}

\begin{figure}
\centering
\includegraphics[width=8.5 cm]{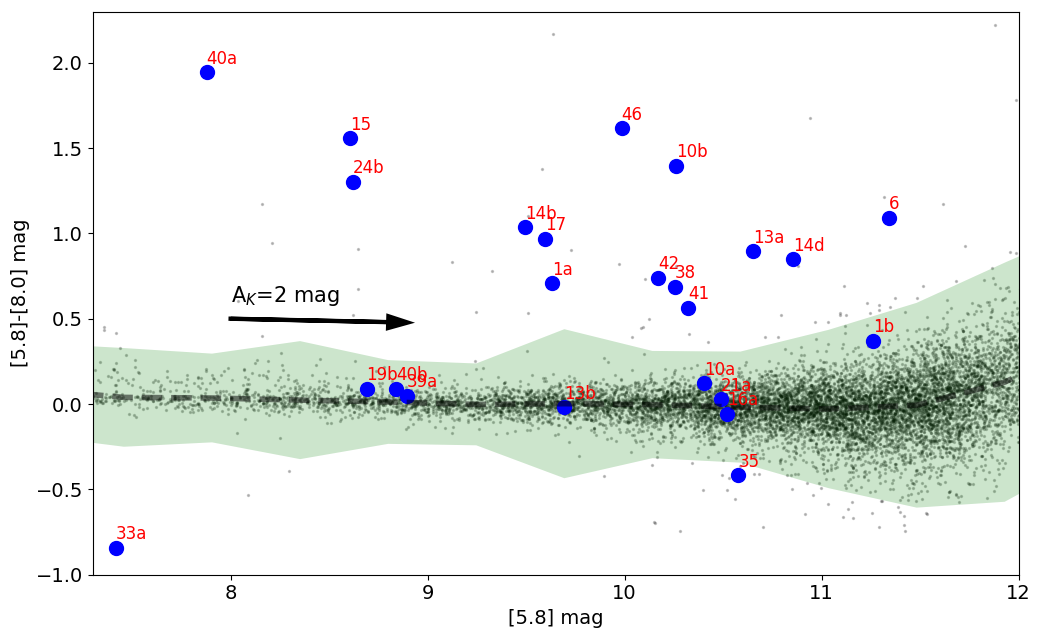}
\vspace{-0.2cm}
\caption{[5.8] -- [8.0] versus [5.8] colour-magnitude diagram for the potential 
outflow driving candidates (blue dots) and sources of the field region (black dots). 
The shaded area represents the $\pm$3$\sigma$ zone around the median [5.8] -- [8.0] colour of the field stars. 
}
\label{iracmd}
\end{figure}

Based on  IRAC colours or fitting models to the observed SEDs or excess X-ray emission, a
number studies have been made in the literature to identify YSOs in the  M17  complex 
\citep[][hereafter literature catalogues]{pov09,pov10,pov13,broos13,bha13}. For identifying potential YSO candidates 
in the vicinity of the jets/knots, we began our search by using the above literature catalogues, however,
these catalogues are incomplete as none of them have incorporated the 70 $\mu$m data and some of them have not used 24 $\mu$m data. 
Moreover,  we found  many bright 24 $\mu$m sources in the vicinity of the jets/knots are either visible 
in only one or two IRAC bands (e.g. see Fig. \ref{fig_mho23}). These sources could be YSOs, but not listed in the literature 
as they failed to pass through the colour combinations or data range used for the YSO identification. 
Since we are interested only in those sources that are close to the jets/knots and, moreover, the
number of  MHOs in our case is reasonably small, a detailed inspection and classification of point
sources around each MHO are possible. Therefore, in addition to the above literature catalogues, we
used a set of colour diagnosis between 3.6 to 70 $\mu$m  (described below) to classify 
all those potential point sources (e.g. see sources 21a, 21b and 21c in Fig. \ref{fig_mho23}) whose possibility as an outflow driving source exists. 
We particularly paid attention to those sources that are detected at 24/70 $\mu$m and 
lies close to the jets/knots. For sources that are visible at $\leq$ 8$\mu$m, we mostly used literature catalogues to infer
their evolutionary status. 

The infrared spectral index ($\alpha$), which is the slope of the SED between 2 and 24 $\mu$m, is used as an  indicator of the YSO 
evolutionary stage \citep{lada87}. In this scheme, $\alpha$ value $>$ 0 represents 
class 0 or I classes (i.e. rising SED sources with strong spherical envelopes), 
$\alpha$ value between 0 to -2 represents 
class II class (i.e. PMS stars with optically thick accreting discs) and  $\alpha$ value between -2 to -3 represents class III class 
(i.e. PMS stars with little or no discs left).  
\citet{rob06} introduced the alternative nomenclature ``stage 0/I, II, III'', 
equivalent to the above classes, but based on the physical properties of the YSOs obtained using radiative transfer models.
Based on the distribution of the young sources of the above classes or stages on the IRAC and MIPS colour-colour diagrams,
several colour schemes have been developed in the literature to classify YSOs \citep[e.g.][]{all04,gut09,rob07,rub11}. However, we note that
ambiguity is an inherent property of any YSO taxonomy, whether based upon colours, spectral indices or physical properties,
and various schemes can alter classification by 20 to 30 per cent \citep[e.g.][]{cra08,hei15,car16}.
Keeping this caveat in mind, for simplicity, here we used the term ``class'' to refer
either the physical stage or observational class of a YSO and adopt the following colour schemes to classify the potential sources into 
various YSO classes:

\begin{enumerate}
\item For those potential sources that are detected in both the IRAC and MIPS bands, 
we used the [3.6] -- [5.8] versus [8.0] -- [24] colour-colour scheme \citep{rob07} for classification. The distribution of such
sources in the [3.6] -- [5.8] versus [8.0] -- [24] colour-colour space is shown in Fig. \ref{iraccc}a. 
As can be seen, the majority of these sources are class 0/I-type YSOs. 
Though high extinction can affect the above classifications, the  effect should be minimal in our case. For example, even if
we deredden the sources with a foreground extinction of \ak $\sim$ 3 mag (indicated in the figure), 
most of our candidates will still remain in the zone of class 0/I.   

\item We observed that a few potential point sources in the vicinity of jets/knots are invisible at 3.6 $\mu$m or 4.5 $\mu$m, 
yet are detected at $\geq$ $5.8$ $\mu$m (possibly due to  extreme visual extinction), while a few 
other point sources have no detection at 5.8 and/or 8.0 $\mu$m band (likely due to low sensitivity of these bands) 
but are detected at $\geq$ 24 $\mu$m.  To classify the sources of the former category, we looked for a correlation between [8] -- [24] colour 
and $\alpha$ using the \citet{dun15}'s YSO sample (shown in Fig. \ref{iraccc}b). \citet{dun15} compiled spectral energy distributions for 2966 YSOs 
and tabulate the infrared spectral index, bolometric luminosity, and bolometric temperature for each YSOs.  
As can be seen from  Fig. \ref{iraccc}b, despite a significant scatter,
a strong correlation between [8] -- [24] colour and $\alpha$  is clearly evident in the figure. From this correlation, we infer 
that sources with [8.0] -- [24] mag $>$ 3.9 and 3.9 $>$ [8.0] -- [24] mag $<$ 1.8 are  respectively, 
the class 0/I and  class II spectral sources. To classify the sources of the latter category, following \cite{gut09} suggestion, 
we considered them as class 0/I YSOs, if  they have  [X] -- [24] colour $>$ 4.5 mag, where [X] is the photometry 
in any of the first two IRAC bands. 

\item We also observed that in the vicinity of a few MHOs, some of the point sources are either invisible or very faint 
in 24 $\mu$m but significantly bright in 70 $\mu$m (e.g. the source 21b in Fig. \ref{fig_mho23}). They could be
deeply embedded YSOs of the complex. In such cases, 
we used the 24 and 70 $\mu$m color combinations for classification. 
The advantage of using 70 $\mu$m is that it is  least sensitive to circumstellar 
extinction and geometry of the disc \citep{dunham06} and proven as an important wavelength 
to identify early class 0 sources. For example, based on {\it Herschel} 
70 $\mu$m observations, \citet{stutz13} identified 18 sources in the Orion complex that are visible at $\geq$ 70 $\mu$m.
They named these sources `PACS Bright Red sources (PBRs)'. 
Comparing the SEDs of the PBRs  with the radiative transfer models they infer that PBRs are very early class 0 objects,
when the envelope is massive and the protostar still has to accrete most of its mass. They found  that PBRs have very red mid- to far-infrared 
colours, i.e.  $log (\lambda F_\lambda 70)/(\lambda F_\lambda 24)$  $\ge$ 1.65. Their analysis also showed that the well-known 
class 0/I sources of the Orion complex from the {\it Herschel} Orion Protostar Survey (HOPS) have 
$log (\lambda F_\lambda 70)/(\lambda F_\lambda 24)$ $>$ 0.0. 
We followed \citet{stutz13} criteria
(shown in Fig. \ref{iraccc}c) to understand the evolutionary status of 
the bright 70 $\mu$m sources   as well as to confirm the classification of the YSOs (i.e. those sources with 70 $\mu$m detection)
identified in the aforementioned schemes. As can be seen from Fig. \ref{iraccc}c, 
most of the 70 $\mu$m detected sources (blue dots in the figure) have $log (\lambda F_\lambda 70)/(\lambda F_\lambda 24)$ value greater than 
0.0, therefore they are most likely protostars. Among which, three sources are likely PBRs (one PBR
is not shown in the plot owing to its high $log (\lambda F_\lambda 70)/(\lambda F_\lambda 24)$ value).

\item 
For a few MHOs, we observed bright IRAC sources at the centre of bipolar flows with no 24/70 $\mu$m counterparts 
(e.g. the source 21a in Fig. \ref{fig_mho23}). These sources could be low-luminosity class II/III YSOs, so the  possibility of 
missing such YSOs in the aforementioned schemes exists. In such cases, we used literature catalogues, 
as many of them have used near-infrared  and X-ray excess emission characteristics of the point sources 
to classify them as YSOs. It is worth mentioning that X-ray emission is more sensitive to the detection
of class III YSOs while near-infrared excess is more sensitive to the detection of class II YSOs. 
In addition, we also used [5.8] versus [5.8] -- [8.0] colour magnitude diagram (CMD) to examine the possibility of such 
sources as YSOs, because [5.8] -- [8.0] colour is an extinction-free indicator as extinction laws are nearly same at 5.8 and 8.0 $\mu$m 
(\eg A$_{5.8}$/A$_K$=0.40 and A$_{8.0}$/A$_K$=0.41; \cite{wei01}, see also \cite{cha09}).
Thus in the [5.8] versus [5.8] -- [8.0] diagram, infrared-excess sources are expected to appear red in colour, 
whereas the background reddened sources in the direction of M17 would appear nearly colourless and are 
likely to fall in the zone of field stars. To see the distribution of field stars in the CMD, we used 
a  field region located $\sim$ 1 degree south-east  of the \hii region, and is devoid of cold dust emission and has 
no active site of star-formation as per SIMBAD database.
Figure \ref{iracmd} shows the distribution
of the field stars (grey dots) and all the potential IRAC sources including those already identified as YSOs 
in the  aforementioned schemes (blue dots). 
From Fig \ref{iracmd}, we considered a potential IRAC only source as a possible YSO, 
if its [5.8] -- [8.0] colour  is redder than 3$\sigma$ of the median [5.8] -- [8.0] colour of 
the field stars (shown as blue shaded area) at its corresponding 5.8 $\mu$m magnitude.  
Doing so, we found most of the potential IRAC only sources  are likely field stars as most of them are located in the shaded area of the plot and 
none of them have been identified as YSOs in the literature.

In all the figures, the ID numbers represent the IDs assigned to the MHOs (see Table. 1), and the subscripts a, b and c, of a given ID represent 
the multiple potential sources of that ID that are under discussion (e.g. see Fig. \ref{fig_mho23}). All the above potential driving sources 
are marked and discussed in Appendix B, where images of individual MHOs are shown and discussed. 

After classifying all the potential sources into various YSO classes as described above, along with the catalogues of clumps/cores and 
other indicators, we looked  for the most probable driving sources of the MHOs. The summary of our approach is briefly mentioned in 
Section \ref{iden_mho} and the details are discussed in Appendix B when the individual MHOs are discussed. Briefly, we disqualified 12 
nearby sources (these are IDs 1b, 2, 10a, 13b, 16a, 16b, 19b, 21a, 33a, 35, 39a and 40b of Figs. \ref{iraccc} and \ref{iracmd}) 
from our driving source list. Among the rest of the potential sources, we conclude 18 outflows are driven by protostars (class 0/I YSOs), 2 are driven
by evolved sources (class II YSOs) and 4 are dirven by one of the member of a group of YSOs (or small cluster).
\\

\end{enumerate}


\section{MULTIWAVELENGTH IMAGES AND NOTES ON INDIVIDIUAL MHOs}
Here, we present multiwavelength figures around the MHOs and discussion concerning their morphologies, possible driving sources and association 
with dust the continuum clumps/cores, dark-clouds and SiO emission. In all the figures, the images and symbols, respectively,
have the same meaning as in Fig. \ref{fig_mho23} and Fig. \ref{fig_mho09}, except the SCUBA cores which are (if found)
shown in diamond symbols. The figures and related discussion are available as supporting information with the online version of the article. 
We note that the reduced continuum subtracted \htt images used in this work are publicly available at 
http://astro.kent.ac.uk/uwish2/index.html. And all the multiwavelength figures of the MHOs are avilable at the MHO database hosted at 
http://astro.kent.ac.uk/~df/MHCat/. 

\end{document}